\documentclass[11pt,a4paper]{article}
\pdfoutput=1
\usepackage{jheppub}
\usepackage{graphicx}
\usepackage{amsmath}
\usepackage{amssymb}
\usepackage{array}
\usepackage{color}

\arxivnumber{1307.1373}

\title{\boldmath  Supersymmetric $SU(5)$ Grand Unification for a Post Higgs Boson Era}

\author{D.~J.~ Miller}
\author{and A.~P.~Morais}
\affiliation{SUPA, School of Physics and Astronomy, University of Glasgow, Glasgow, G12 8QQ, UK}

\emailAdd{David.J.Miller@Glasgow.ac.uk}
\emailAdd{a.morais@physics.gla.ac.uk}

\abstract{We investigate models of supersymmetric grand unification based on the gauge group $SU(5)$. We consider models with non-universal gaugino masses and confront them with low energy constraints, including the Higgs boson mass and the Dark Matter relic density. We also discuss fine-tuning and show the effect of not including the $\mu$-parameter into fine tuning determinations. With this relaxation, we find viable scenarios with low fine tuning and study some model choices for gaugino mass ratios. We demonstrate that some orbifold inspired models may provide low fine-tuning and the preferred relic abundance of Dark Matter while evading all experimental constraints.  We present benchmarks that should be explored at the LHC and future colliders. }

\begin{document} 
\maketitle
\flushbottom

\setcounter{footnote}{0}

\section{Introduction}

\hspace{5mm} The discovery of a Higgs-like resonance with a mass of around $125\,{\rm GeV}$ by the ATLAS and the CMS Collaborations {\cite{ATLAS:2012ae, Chatrchyan:2012tx}}, may provide the last piece of the Standard Model (SM). So far this resonance appears consistent with SM production and decay, but is also consistent with supersymmetric models in the decoupling limit where the lightest Higgs boson inherits couplings very similar to its SM counterpart. This needn't have been the case; in the Minimal Supersymmetric Standard Model (MSSM) the lightest Higgs boson is predicted to have mass $ \lesssim 135\,{\rm GeV}$~\cite{Carena:2002es}, so the observation of a Higgs boson only $10\%$ heavier could have ruled out minimal supersymmetry. Even  a mass of $125\,{\rm GeV}$ is not trivial to obtain from models of minimal supersymmetry: the tree-level contribution to the mass is required to be less than the mass of the $Z$-boson, $M_Z$, so one must rely on radiative corrections from  top quarks and their scalar partners. This indicates that the top squark and the supersymmetry breaking scale may be rather heavy.

It should not be so surprising then that searches for supersymmetry at the Large Hadron Collider (LHC)~\cite{ATLAS_susy,CMS_susy} have so far been negative. Most of these searches have been in the context of the {\it constrained} Minimal Supersymmetric Standard Model (cMSSM, see \cite{Martin:1997ns} for a review), where the soft supersymmetry breaking masses and trilinear couplings are universal at the scale where the gauge couplings unify. The LHC experiments have ruled out first and second generation squarks lighter than about $1.5\,{\rm TeV}$, whereas the gluino has to be heavier than about $850\,{\rm GeV}$. While these results put pressure on the cMSSM, there is still plenty of room for the discovery of a heavier supersymmetric spectrum at the LHC or at future colliders (for example a Large Hadron electron Collider (LHeC) \cite{AbelleiraFernandez:2012cc},  the International Linear Collider (ILC) \cite{Brau:2012zz} or a Compact Linear Collider (CLiC) \cite{Lebrun:2012hj}). 

However, a heavy supersymmetric spectrum suffers from fine-tuning problems. One of the main motivations of supersymmetry is its solution to the hierarchy problem, where top squark loops cancel the quadratic divergence of the Higgs mass arising from top quark loops. If the top squarks are too heavy, the remaining non-cancelled logarithmic divergence will also require fine-tuning, resulting in a ``little hierarchy problem.'' Furthermore, soft supersymmetry breaking masses appear in the expression for the $Z$-boson mass, so if these masses are large one requires cancellation of large (squared) masses to reproduce the relatively low mass of the $Z$-boson. These fine-tuning issues are far less problematic than the original hierarchy problem of the SM, but they may still be used to gain additional insight into new physics models.

Despite the great success of the SM, the origin of its gauge structure, $SU(3) \times SU(2) \times U(1)$, is still an unanswered question. Is this symmetry a remnant of some larger simple group that is spontaneously broken at a high scale by some Higgs-like mechanism~\cite{Georgi:1974sy}, or does it derive from some other physics, such as higher dimensional operators or extra dimensions (see for example~\cite{Wang:2011eh})? A Grand Unified Theory (GUT) is strongly motivated by the running of gauge couplings, which within supersymmetric scenarios converge to a common value at a scale of about $M_{GUT} \approx 2 \times  10^{16}\,{\rm GeV}$~\cite{Dimopoulos:1981yj}. The most popular candidates of a unified gauge symmetry are the $SU(5)$, $SO(10)$ and $E_6$ groups~\cite{Ramond:1979py}, and these gauge structures may constrain the soft supersymmetry breaking parameters at the high scale differently from the cMSSM. 

Although the Higgs boson mass indicates that top squarks are most likely rather heavy, we note that searches for third generation squarks~\cite{ATLAS_3g} remain relatively weak in comparison with the first two generations. This leaves room for scenarios with non-universal scalar masses  across the generations that may have very heavy first and second generation squarks but relatively light top squarks. As in the SM, hierarchies of a few order of magnitude between generations should not be surprising, though vastly different GUT scale scalar masses would be difficult to generate using the same mechanism.

Similarly, the gaugino fermions provide another source of possible non-universality (see e.g.\ \cite{Abe:2007kf,Gogoladze:2009bd,Horton:2009ed,Antusch:2012gv,Caron:2012sf}). Gauginos are embedded in the adjoint representation of the GUT gauge group and their masses may be generated by hidden-sector chiral superfields in the gauge-kinetic function developing expectation values. If the chiral superfields are a non-singlet representation of the gauge symmetry, we may also have non-universal GUT scale gaugino masses. 

Scenarios with non-universal masses are important within the context of dark matter (DM) studies. Universal gaugino masses typically predict bino dominated neutralinos as the Lightest Supersymmetric Particle (LSP) with a relic density a few orders of magnitude beyond the upper limit set by the Wilkinson Microwave Anisotropy Probe (WMAP)~\cite{Hinshaw:2012fq} and Planck~\cite{Ade:2013xsa} satellites. 
On the other hand, non-universal gauginos are compatible with Higgsino and wino dominated DM, and for a judicious choice of masses, the neutralino relic density can be within or below the WMAP and Planck bounds. In this paper,  non-universality of gaugino masses in a $SU(5)$ GUT is explored and the regions of the parameter space that favour a DM candidate with acceptable relic density are discussed. Our analysis is not only restricted to the gaugino sector; we also explore possible non-universalities arising from the $SU(5)$ boundary conditions, assuming that the GUT embedding should leave its signature in the sfermion masses, as well as in the soft trilinear couplings. 

Although the extra super-heavy gauge bosons belonging to the off-diagonal elements of the adjoint representation may mediate proton decay, the supersymmetric GUT scale, which is few orders of magnitude higher than that of a non-supersymmetric scenario, is sufficient to suppress the baryon number violating interactions. However,  in general, Grand Unified Theories in which the GUT gauge group breaks directly to the Standard Model suffer from a doublet-triplet splitting problem \cite{Sakai:1981gr}. The presence of coloured fields in (enlarged $SU(5)$) Higgs multiplets would not only spoil the unification of the gauge couplings, but also mediate fast proton decay through baryon number violating interactions \cite{proton_decay}. The presence of higher dimensional operators may also be problematic for proton stability. However, it has also been pointed out that proton decay may not be an issue for minimal $SU(5)$ after all due to uncertainties in either sfermion masses and mixings or the triplet mass~\cite{Bajc:2002bv}. In the present work, we will implicitly assume that these issues are either not problematic or are solved by some unknown mechanism at the GUT scale, such as embedding the model in higher dimensions \cite{higher_dim}. 

The remaining sections of this paper are organized as follows. In Sec.~\ref{sec:su5pars} we will outline the $SU(5)$ GUT model and clarify our parameters. In Sec.~\ref{sec:scans} we will outline out implementation of experimental and stability constraints and discuss our philosophy regarding fine-tuning. We will provide an initial analysis of universal gaugino masses in Sec.~\ref{sec:ugm} before going on to investigate non-universal gaugino masses in Sec.~\ref{sec:nugm}. This will identify specific desirable gaugino mass ratios and we investigate particular choices (inspired by high scale models) in Sec.~\ref{sec:sfgmr}. Finally we will present some benchmark scenarios in Sec.~\ref{sec:BP} and conclude in Sec.~\ref{sec:conc}.

%%%%%%%%%%%%%%%%%
\section{The $SU(5)$ GUT Model}\label{sec:su5pars}

We consider a $SU(5)$ GUT model with the superpotential given by
\begin{equation}
W_{SU(5)} = \varepsilon_{\alpha \beta \gamma \rho \sigma}\left(y_{\mathbf{5^{\prime}}}\right)_{ij}\mathbf{10}^{\alpha \beta}_i \mathbf{10}^{\gamma \rho}_j \mathbf{5^{\prime}}^{\sigma} + \left(y_{\mathbf{\overline{5}^{\prime}}}\right)_{ij}\mathbf{10}^{\alpha \beta}_i \mathbf{\overline{5}}_{j \alpha} \mathbf{\overline{5}^{\prime}}_{\beta} + \mu \mathbf{\overline{5}^{\prime}}_{\alpha} \mathbf{5^{\prime}}^{\alpha} + W_{\mathbf{X_{\mathbf{R}}}}.
\label{eq:W5}
\end{equation}
Here, $W_{\mathbf{X_{\mathbf{R}}}}$ is the part of the superpotential that involves the chiral superfields $X_{\mathbf{R}}$, belonging to a $SU(5)$ symmetric representation $\mathbf{R}$, contained in the product of two adjoint representations, $\mathbf{24} \times \mathbf{24}$, and whose scalar components are responsible for the breaking of the GUT symmetry at the high scale.  Greek letters are $SU(5)$ indices, Roman letters are generation indices and $\varepsilon_{\alpha \beta \gamma \rho \sigma}$ is the five dimensional generalization of the Levi-Civita symbol. The left-handed quark doublet $\hat{Q}_L$, right-handed up-quark $\hat{u}^{\dagger}_R$ and right-handed charged lepton $\hat{e}^{\dagger}_R$ superfields are embedded in the $\mathbf{10}$ representation, while the left-handed lepton doublet $\hat{L}_L$ and right-handed down-quark $\hat{d}^{\dagger}_R$ superfields are in the $\overline{\mathbf{5}}$ representation. The Higgs superfields $\hat{H}_u$ and $\hat{H}_d$ are in the $\mathbf{5^{\prime}}$ and $\mathbf{\overline{5}^{\prime}}$ representations respectively. These are indeed the surviving fields after the breaking of the GUT symmetry to the SM gauge group $G_{SM}$, where we have assumed that the doublet-triplet splitting problem is solved by some unknown mechanism at the high scale, as discussed in the introduction. 
Since supersymmetry has not been observed at low energies, we must of course break it, possibly in a hidden sector (see, for example,  \cite{Intriligator:2007zz}), which manifests as soft  supersymmetry-breaking terms \cite{Chung:2003fi} in the Lagrangian.

%%%%%%%%%%%%%%%%%
\subsection{Soft Scalar Masses}

For simplicity we will here assume {\it gravity mediated} or {\it Planck-suppressed} supersymmetry breaking. We have an effective theory below the Planck scale containing higher dimensional operators suppressed by the Planck mass, which may, in a string theoretic approach, arise from the compactification of extra dimensions. These operators couple the fields of the hidden sector to the fields of the visible sector, and the scalar masses may arise from the dimension-6 operators
\begin{equation}
-\mathcal{L}_{dim-6} = \frac{\kappa^i_{~j}}{M^2_{P}} |F_X|^2  \tilde \phi_i \tilde \phi^{* j},
\label{eq:L-6I}
\end{equation}
where $F_X$ is an F-term of a hidden sector superfield $\hat X$, $\tilde{\phi}$ is the scalar component of a visible sector superfield $\hat \Phi$ with mass $m_{\tilde{\phi}}$, $\kappa^i_{\:j}$ a coupling and $M_P$ is the Planck mass.
If the F-term $F_X$ has a non-vanishing expectation value, the scalar masses are
\begin{equation}
\left( m^2_{\tilde{\phi}} \right)^i_{~j} \equiv
\frac{\kappa^i_{~j}}{M^2_{P}} \left| \langle F_X \rangle \right|^2.
\label{eq:L-6II}
\end{equation}

In the MSSM, the part of the Lagrangian that includes the Higgs and sfermion soft masses is given by
\begin{eqnarray}
-\mathcal{L}_{mass} &=&
 m^2_{H_d}\left|H_d\right|^2 +  m^2_{H_u}\left|H_u\right|^2 + \tilde{Q}^{~\alpha x}_{Li} \left( m^2_{\tilde{Q}_L} \right)^i_{~j} \tilde{Q}^{\ast~j}_{L \alpha x} + \tilde{L}^{~\alpha}_{L i} \left( m^2_{\tilde{L}_L} \right)^i_{~j}\tilde{L}^{\ast~j}_{L \alpha}\nonumber\\
 &+& \tilde{u}^{\ast}_{R i x} \left( m^2_{\tilde{u}_R} \right)^i_{~j}\tilde{u}^{~j}_{R x} + \tilde{d}^{\ast}_{R i x} \left( m^2_{\tilde{d}_R} \right)^i_{~j}\tilde{d}^{~j}_{R x} + \tilde{e}^{\ast}_{R i} \left( m^2_{\tilde{e}_R} \right)^i_{~j}\tilde{e}^{~j}_{R },
\label{eq:Lmssm}
\end{eqnarray}
where the squared soft masses run according to the Renormalization Group Equations (RGE) \cite{Martin:1993zk}. In this Lagrangian, $i,j=1,2,3$ are again generation indices, but now $\alpha = 1,2$  are weak isospin and $x = 1,2,3$ is a colour index.
For a standard $SU(5)$ GUT, when the unified symmetry is broken to $G_{SM}$, the sfermions, which are embedded in $\mathbf{10}$ and $\mathbf{\overline{5}}$ dimensional representations, take soft masses $m_{\mathbf{10}}$ or $m_{\mathbf{\overline{5}}}$. Furthermore, we allow an hierarchy between the third generation and the first two generations, but keep the first two generations degenerate in order to avoid dangerous Flavour-Changing Neutral-Currents (FCNC)~\cite{Baer:2004xx}. Therefore, this model has two extra parameters, $K_{\mathbf{\overline{5}}} > 0$ and $K_{\mathbf{10}} > 0$, which account for the third generation's non-universality at the GUT scale. For the Higgs sector, the masses of the doublets that couple to the up-type quarks and down type quarks take the high scale values $m_{\mathbf{5}^{\prime}}$ and $m_{\mathbf{\overline{5}^{\prime}}}$ respectively. Our boundary conditions for the scalar soft masses at the GUT scale are then given by:
\begin{eqnarray}
m^2_{Q_{ ij}}\left( 0 \right) \:\: =\:\: m^2_{u_{ij}}\left( 0 \right)  \:\: =\:\:   m^2_{e_{ij}}\left( 0 \right) &=& 
\begin{pmatrix}
K_{\mathbf{10}} & 0 & 0\\
0 & K_{\mathbf{10}} & 0\\
0 & 0 & 1
\end{pmatrix} m^2_{\mathbf{10}},
\label{eq:10scalarBC}\\[2mm]
m^2_{L_{ ij}}\left( 0 \right)  \:\: =\:\:  m^2_{d_{ij}}\left( 0 \right) &=& 
\begin{pmatrix}
K_{\mathbf{\overline{5}}} & 0 & 0\\
0 & K_{\mathbf{\overline{5}}} & 0\\
0 & 0 & 1
\end{pmatrix}m^2_{\mathbf{\overline{5}}},
\label{eq:5scalarBC}\\[2mm]
m^2_{H_u}\left( 0 \right) &=& m^2_{\mathbf{5}^{\prime}}, \label{eq:HuBC}\\
m^2_{H_d}\left( 0 \right) &=& m^2_{\mathbf{\overline{5}}^{\prime}}. \label{eq:HdBC}
\end{eqnarray}
In the above, the RGEs are parameterized by $t \equiv \log(Q/Q_0)$, where $Q$ the energy scale of interest and $Q_0$ is the unification scale.

To accompany the $\mu$-term in Eq.(\ref{eq:W5}) we also have a soft scalar mass term of the form $\varepsilon_{\alpha \beta} \left[ b H^{\alpha}_d H^{\beta}_u +h.c. \right]$ where $\varepsilon_{\alpha \beta}$ is an antisymmetric tensor with $\varepsilon_{12} = 1$. However, $b$ is determined from the electroweak symmetry breaking (EWSB) condition\footnote{In our analysis we use the two-loop generalisation of Eq.~(\ref{eq:EWSB}).}
\begin{eqnarray}
b = \frac{\sin 2 \beta}{2} \left( m^2_{H_u} +  m^2_{H_d} + 2\mu^2 \right),
\label{eq:EWSB}
\end{eqnarray}
so, unlike the other soft supersymmetry breaking parameters, it is not a high scale input for our analysis.

%%%%%%%%%%%%%%%%%
\subsection{Soft Trilinear Couplings}

Soft trilinear terms may arise from dimension five operators of the form
\begin{equation}
-\mathcal{L}_{dim-5} = \frac{\eta^{i j k}}{M_{P}} F_X \tilde \phi_i \tilde \phi_j \tilde \phi_k.
\label{eq:L-5I}
\end{equation}
When the F-terms of $\hat X$ develop an expectation value, such terms generate the scalar trilinear couplings
\begin{equation}
a^{ijk}\equiv
\frac{\eta^{i j k}}{M_{P}} \langle F_X \rangle.
\label{eq:L-5II}
\end{equation}

The explicit soft susy-breaking terms that contain scalar trilinear couplings are given by
\begin{eqnarray}
-\mathcal{L}_{trilinear} = \varepsilon_{a b} \left[ a_{u ij}H^a_u \tilde{u}^{~x}_{R i}\tilde{Q}^{b}_{L j x} - a_{d ij}H^a_d \tilde{d}^{~x}_{R i}\tilde{Q}^{b}_{L j x} - a_{e ij}H^a_d \tilde{e}_{R i}\tilde{L}^{b}_{L j} + h.c. \right],
\label{eq:Ltmssm}
\end{eqnarray}
where the indices have the same meaning as in Eq.~(\ref{eq:Lmssm}). It is usual to define the trilinear couplings in terms of the Yukawa couplings as $\left(a_{u,d,e}\right)_{ij} = \left(y_{u,d,e}\right)_{ij}\left(A_{u,d,e}\right)_{ij}$. Since the first and second generation Yukawa couplings are very small, we only consider contributions from the third generation trilinears and Yukawa couplings. The $\left(a_{u,d,e}\right)_{ij}$ are then effectively diagonal with only one non-zero entry each \mbox{$\left(a_u\right)_{33} \equiv a_t$},  \mbox{$\left(a_d\right)_{33} \equiv a_b$} and $\left(a_e\right)_{33} \equiv a_\tau$, and we impose the boundary conditions
\begin{eqnarray}
a_{t}\left( 0 \right) &=& a_{\mathbf{5}^{\prime}}, \label{eq:AuBC}\\
a_b\left( 0 \right) = a_{\tau}\left( 0 \right) &=& a_{\mathbf{\overline{5}}^{\prime}}. \label{eq:AdeBC}
\end{eqnarray}
Since $\hat t_R^\dagger$ is in a different $SU(5)$ multiplet from $\hat b_R^\dagger$ and $\hat \tau_R^\dagger$ we make no attempt to unify the top Yukawa coupling with those of the bottom or $\tau$ at the high scale. 

%%%%%%%%%%%%%%%%%
\subsection{Gaugino Masses} \label{subsec:gugino}

Gaugino masses may arise from a gauge-kinetic term of the form \cite{Cremmer:1978iv,Ellis:1985jn,Cerdeno:1998hs,Bhattacharya:2009wv}
\begin{eqnarray}
\mathcal{L}_{g-k} &=& \int d^2 \theta f_{\alpha \beta}\left( \hat X_i \right) \hat W^{a \alpha} \hat W^{\beta}_a + h.c. \nonumber \\ &=& - \frac{1}{4} Re f_{\alpha \beta}  F^{\alpha}_{\mu \nu} F^{\beta \mu \nu} + \frac{1}{4} e^{-G/2} \frac{\partial f^{*}_{\alpha \beta}}{\partial \varphi^{j *}} \left( G^{-1} \right)^j_k G^k \tilde \lambda^{\alpha} \tilde \lambda^{\beta} + \cdots
\label{eq:Lgk}
\end{eqnarray}
$\hat W^{a \alpha}$ is the gauge field strength superfield, $F^{\alpha}_{\mu \nu}$ is the field strength tensor and $\tilde \lambda^{\alpha}$ is a gaugino fermion; $\alpha$ and $\beta$ are gauge indices, $a$ is a spinor index,  and as usual $\mu$ and $\nu$ are Lorentz indices. $\hat X_i$ are again the hidden sector superfields but now we include an index $i$ in recognition that there may be more than one.  The gauge-kinetic function $f_{\alpha \beta}\left( \hat X_i \right)$ is an analytic function of the $\hat X_i$ superfields transforming as a symmetric product of two adjoint $\mathbf{24}$ representations of $SU(5)$ so that the the Lagrangian is gauge invariant. $G\left( \hat X_i,\hat X^*_i \right)$ is a real function \mbox{$G = K + \log \lvert W \rvert^2$} where K is the K\"{a}hler potential and $W$ the superpotential. $G^k \equiv \partial G / \partial \varphi_k$ and $G^j_k \equiv \partial^2 G / \partial \varphi_j \partial \varphi^{k *}$ with $\left( G^{-1} \right)^i_k G^k_j =\delta^i_j$, where $\varphi_i$ is the scalar component of $\hat X_i$. When an F-term $F_X$ develops an expectation value, it spontaneously breaks supersymmetry and enters Eq.~(\ref{eq:Lgk}) by identifying
\begin{eqnarray}
F^j _X= \frac{1}{2} e^{-G/2}\left[ \left( G^{-1} \right)^j_k G^k \right] ,
\label{eq:Fterm}
\end{eqnarray}
generating a gaugino mass term of the form
\begin{eqnarray}
\frac{1}{2} \langle F^j_X \rangle \left\langle \frac{\partial f^{*}_{\alpha \beta}}{\partial \varphi^{j *}} \right\rangle \tilde \lambda^{\alpha} \tilde \lambda^{\beta}.
\label{eq:LM}
\end{eqnarray}
The representations of the $\hat X_i$ are unknown, but we may expand the gauge-kinetic function in terms of singlet $\hat X^S$ and non-singlet $\hat X^N$ superfields
\begin{eqnarray}
f_{\alpha \beta}\left( \hat X^i \right) = f_0\left( \hat X^S \right) \delta_{\alpha \beta} + \sum_{N} f_N\left( \hat X^S \right) \frac{\hat X^N_{\alpha \beta}}{M_P} + {\cal O}\left( 1/M^2_P \right),
\label{eq:f}
\end{eqnarray}
where $f_0$ and $f_{N}$ are functions of the singlet fields only. When this is inserted into the first term in the right-hand-side of Eq.~(\ref{eq:Lgk}) we have additional five-dimensional operators 
which generate an extra contribution to the canonical gauge-kinetic terms $-\frac{1}{4} F^{\alpha}_{\mu \nu} F^{\alpha \mu \nu}$.
It has been shown \cite{Ellis:1985jn,Bhattacharya:2009wv, Hill:1983xh, Chakrabortty:2008zk} that such operators do not spoil the unification of the gauge couplings both at one-loop and two-loop level and we may return to the canonical form by a rescaling of the superfields.

After this rescaling, the gaugino mass terms take the form
\begin{eqnarray}
\frac{1}{2} \frac{\langle F^j_X \rangle}{\langle Re f_{\alpha \beta} \rangle} \left\langle \frac{\partial f^{*}_{\alpha \beta}}{\partial \varphi^{j *}} \right\rangle\tilde \lambda^{\alpha} \tilde \lambda^{\beta}.
\label{eq:LM}
\end{eqnarray}
The coefficient is a representation (or a combination of representations) belonging to the product $\left(\mathbf{24} \times \mathbf{24}\right)_{symm} = \mathbf{1} + \mathbf{24} + \mathbf{75} + \mathbf{200} $. If it is a singlet only the first term of Eq.~(\ref{eq:f}) is relevant and we have a universal gaugino mass for the SM gauge groups,
\begin{eqnarray}
M_{1/2} = \frac{\langle F^j_X \rangle}{\langle Re f_{0} \rangle} \left\langle \frac{\partial f^{*}_{0}}{\partial \varphi^{j *}} \right\rangle.
\label{eq:universal}
\end{eqnarray}
However, this needn't be the case and the coefficient may be in a more non-trivial representation (or a combination of them), resulting in $SU(3)$, $SU(2)$ and $U(1)$ gauginos that have non-universal masses at the high scale.
The effective soft gaugino-mass terms are then
\begin{eqnarray}
\frac{1}{2} \left[ M_1 \tilde{\lambda}_1\tilde{\lambda}_1 + M_2 \tilde{\lambda}_2\tilde{\lambda}_2 +  M_3 \tilde{\lambda}_3\tilde{\lambda}_3  \right].
\label{eq:LM12}
\end{eqnarray}
We will therefore examine two distinct sets of boundary conditions at the GUT scale:
\begin{itemize}
\item[I.] universal gaugino masses: $M_1 = M_2 = M_3 \equiv M_{1/2}$,
\item[II.] non-universal gaugino masses: $M_1/\rho_1 = M_2/\rho_2 = M_3 \equiv M_{1/2}$,
\end{itemize}
where $\rho_1$ and $\rho_2$ are new parameters we introduce to quantify the non-universality.

%%%%%%%%%%%%%%%%%
\subsection{Summary of the Parameter Space}

In addition to the usual SM parameters, our $SU(5)$ model is described by eleven high scale parameters, $m_{\overline{5}}$, $K_{\overline{5}}$, $m_{10}$, $K_{10}$,  $M_{1/2}$,  $\rho_1$, $\rho_2$, $ m_{\overline{5}^{\prime}}$, $m_{5^{\prime}}$, $a_{\overline{5}^{\prime}}$, $ a_{5^{\prime}}$, as well as $\tan \beta$ and the sign of $\mu$. The value of $\mu^2$ is fixed by the Z boson mass as usual.

%%%%%%%%%%%%%%%%%
\section{Constraints on the Particle Spectrum}
\label{sec:scans}

The next step is to use the RGEs to evolve the soft masses and couplings down to the electroweak scale, where the particle spectrum may be confronted with the various experimental constraints and possible fine-tunings examined. We perform this running using SOFTSUSY 3.3.0 \cite{Allanach:2001kg}, starting from the boundary conditions described in section~\ref{sec:su5pars}.

We allow the third generation GUT scale scalar masses, $m_{\overline{5}}^{(3)}$ and $m_{10}^{(3)}$ to lie between zero and $3.5\,{\rm TeV}$ and then choose $K_{10}$, $K_{\overline{5}}$ between zero and $10$ to give the first and second generation scalar masses. The high scale masses of the Higgs multiplets, $m_{\overline{5}^{\prime}}$ and $m_{5^{\prime}}$ are constrained to be less then $4\,{\rm TeV}$. We require $M_3$ to be less than $2\,{\rm TeV}$; if examining scenarios with  universal gaugino masses, this also sets $M_1$ and $M_2$, but if examining non-universal gauginos, we also vary $\rho_{1,2}$ between $\pm 15$.
Finally the trilinear couplings, $a_{\overline{5}^{\prime}}$ and $a_{5^{\prime}}$, are allowed to vary between $\pm 10\,{\rm TeV}$, and our only (non-SM) low energy input $\tan \beta$ is constrained to lie in the range $1\--60$.

We generate scenario points randomly within these ranges, separately for universal and non-universal gaugino masses. Although the input parameters for the generated scenarios are evenly distributed within their allowed ranges, we make no attempt to ascribe a significance to this distribution. Since the dynamics of the hidden sector are unknown to us, we assign no prior probability for the distribution of input parameters in theory space,  and do not perform a Bayesian analysis of the low energy scenarios. The random inputs are then only an attempt to fill parameter space with possible scenarios and their density holds no significance. This is a rather different approach from some analyses in the literature~\cite{Baltz:2004aw} where theoretical priors are assigned.

%%%%%%%%%%%%%%%%%
\subsection{Experimental Constraints}
\label{sec:expc}

Each scenario must be confronted by experiment. Our first such constraints are the LHC direct searches for supersymmetry from ATLAS~\cite{ATLAS_susy} and CMS~\cite{CMS_susy}. These limits are rather non-trivial surfaces in parameter space (for example, the limit on the gluino mass is dependent on the squark masses) but here, in the interest of simplicity, we make simple, though more conservative cuts on individual masses. In particular, we require the first and second generation squarks to have masses greater than $1.4\,{\rm TeV}$, the gluino to be heavier than $800\,{\rm GeV}$ and the lightest chargino heavier than $103.5\,{\rm GeV}$. We do not explicitly constrain the third generation squarks since we find scenarios that violate the appropriate searches~\cite{ATLAS_3g} are already ruled out by other experimental constraints. The only other direct cut we make is for the direct detection of Dark Matter; we use 
micrOMEGAS 2.4.5 \cite{Belanger:2004yn} to calculate the spin independent cross section for the scattering of Weakly Interacting Massive Particles (WIMPs) and nucleons, $\sigma^{NW}_{SI}$, and compare with the 2$\sigma$ bounds set by XENON100~\cite{Aprile:2012nq}.

We also confront our model with the newly measured Higgs boson mass as well as the Dark Matter relic density, and bounds on new physics from $b \rightarrow s \gamma$,  $B_s \rightarrow \mu^+ \mu^-$, $B \rightarrow \tau \nu_{\tau}$ and the muon anomalous magnetic moment $a_{\mu}$. For all of these, except for the Higgs boson mass, we again use micrOMEGAS to calculate their values for our scenarios and assume a $10\%$ theoretical error. For each of these measurements we compare our prediction with experiment and determine the probability of the given deviation assuming Gaussian errors. We then combine the individual probabilities into a total probability $P_{\rm tot} = P_{m_h} \cdot P_{\Omega_c h} \cdot P_{b \rightarrow s \gamma} \cdot P_{\mathcal{R}_{\tau \nu_{\tau}}} \cdot P_{B_s \rightarrow \mu \mu} \cdot P_{a_{\mu}}$ and require that this is never smaller than $10^{-3}$. This excludes scenarios with multiple predictions close to their $\pm 2 \sigma$ bound, that would otherwise be accepted by imposing the contraints on a one-by-one basis.

For the Higgs boson mass, we use the ATLAS~\cite{ATLAS:2012ae} and CMS~\cite{Chatrchyan:2012tx} values $126 \pm 0.8\,{\rm GeV}$ and \mbox{$125.3 \pm 0.9\,{\rm GeV}$} respectively. We combine these together and add a $\pm 2\,{\rm GeV}$ theoretical uncertainty in quadrature. This theoretical uncertainty was estimated by the mass difference for the light CP-even Higgs obtained with SOFTSUSY and SUSPECT~\cite{Djouadi:2002ze}, as reported in~\cite{Arbey:2012dq}. This gives (1$\sigma$) uncertainty on our output Higgs boson mass of $125.7 \pm 2.1\,{\rm GeV}$.

Constraints on $b \rightarrow s \gamma$ were taken from the Heavy Flavour Averaging Group \cite{Amhis:2012bh}, who report a  measured value for the branching ratio ${\rm Br} \left(b \rightarrow s \gamma \right) = \left( 355 \pm 24 \pm 9 \right) \times 10^{-6}$. Combining this with the theoretical error provides bounds of \mbox{${\rm Br} \left(b \rightarrow s \gamma \right) = \left( 355 \pm 43.8\right) \times 10^{-6}$}.

First evidence of the decay $B_S \rightarrow \mu^{+} \mu^{-}$ was recently observed by LHCb~ \cite{:2012ct}. A fit to data leads to the decay branching ratio ${\rm Br} \left(B_S \rightarrow \mu^{+} \mu^{-} \right) = \left( 3.2^{+1.5}_{-1.2}\times 10^{-9} \right)$.
These errors are still sufficiently large that the theoretical uncertainty leaves them unchanged.

The latest Belle and BaBar results for the purely leptonic $B \rightarrow \tau \nu_{\tau}$ decay \cite{Adachi:2012mm}, measured the branching ratio ${\rm Br} \left(B \rightarrow \tau \nu_{\tau} \right) = \left( 1.12 \pm 0.22\right) \times 10^{-4}$, which can be compared with the SM prediction of $\left( 0.79 \pm 0.23\right) \times 10^{-4}$~\cite{Biancofiore:2013ki}. MicrOMEGAS outputs the ratio of the predicted branching ratio with that of the SM, $\mathcal{R}_{\tau \nu_{\tau}}$. Again combining with a $10\%$ theoretical uncertainty we find that this output should be constrained by $\mathcal{R}_{\tau \nu_{\tau}}  = 1.42 \pm 0.70$.

The anomalous magnetic moment $a_{\mu} = \left( g-2 \right)_{\mu}/2$ has been determined at BNL \cite{Bennett:2006fi} to be $a_{\mu}(exp) = (11~659~208.9 \pm 6.3) \times 10^{-10}$, which may be compared to the SM prediction~\cite{Davier:2009zi} $a_{\mu}(SM) = (11~659~183.4 \pm 4.9) \times 10^{-10}$. This 3-4$\sigma$ tension of (SM) theory and experiment could be a hint for physics beyond the SM, and may be attributed to supersymmetric contributions~\cite{vonWeitershausen:2010zr}, but it is also possible that some other additional cause is responsible for some or all of the deviation. In this study, we only require  that the supersymmetric contribution is not too large. We calculate the extra contribution arising from our model and compare it with \mbox{$\Delta a_{\mu}(exp - SM) = (25.5 \pm 8.0) \times 10^{-10}$}: if the additional contribution is less than this we set $P_{a_\mu}=1$ for this scenario; but if it is more we use the uncertainty to quantify $P_{a_\mu}$ as described above. 

Finally we turn to the relic abundance of Dark Matter. The cosmological parameters of the nine year WMAP observations were recently published in \cite{Hinshaw:2012fq}, where the fit to the cold Dark Matter relic density, $\Omega_c h^2$, provides a value of $0.1157 \pm 0.0023$. We estimate a $10 \%$ theoretical uncertainty arising from the LSP mass difference calculated with SOFTSUSY and micrOMEGAS and add this in quadrature with the experimental fit standard deviation. The resulting bounds for our micrOMEGAS relic density output are $\Omega_c h^2 = 0.1157 \pm 0.0118$. However, for the purposes of exclusion we only  include the probability  $P_{\Omega_c h}$ if the relic density is too high. Scenarios with values below $\Omega_c h^2 = 0.1157$ are accepted,  but we then use $P_{\Omega_c h}$ in the usual way to determine if this mechanism provides the ``preferred'' relic density or too little. Scenarios with too little are kept because there may be some other contribution to Dark Matter such as an axion from a broken global $U(1)$ symmetry~\cite{Peccei:1977hh} (which may also provide a solution to the strong CP problem~\cite{Cheng:1987gp}).

%%%%%%%%%%%%%%%%%
\subsection{Fine-tuning}

One of the original motivations for low energy supersymmetry was a solution to the fine-tuning (hierarchy) problem of the Higgs bosons mass, so it is sensible to also examine the fine-tuning of our scenarios. Of particular interest here is the fine-tuning of the $Z$-boson mass with respect to the input parameters. We use the measure of fine tuning introduced by Barbieri and Giudice~\cite{Barbieri:1987fn}, for which the {\it partial} fine-tuning is
\begin{eqnarray}
\Delta_{\mathcal{P}_i} = \left\lvert \frac{\mathcal{P}_i}{M^2_Z} \frac{\partial M^2_Z}{\partial \mathcal{P}_i} \right\rvert,
\label{eq:FTdef}
\end{eqnarray}
where $\left\{\mathcal{P}_i\right\}$ is the set of input parameters. The fine-tuning of a specific scenario is the maximum of the partial fine tunings,
\begin{eqnarray}
\Delta = {\rm max}\left\{ \Delta_{\mathcal{P}_i} \right\}.
\label{eq:FT}
\end{eqnarray}
For an alternative measure of fine-tuning see~\cite{Athron:2007ry}.

At tree-level\footnote{This tree-level expression is appropriate at the scale $M_S = \sqrt{m_{\tilde t_1}m_{\tilde t_2}}$ where radiative corrections are minimal~\cite{Gamberini:1989jw}.} the $Z$-boson mass is given by
\begin{eqnarray}
M_Z^2 = -2 \left( m_{H_u}^2+\left| \mu \right|^2 \right) + \frac{2}{\tan^2 \beta} \left( m_{H_d}^2-m_{H_u}^2 \right) + {\cal O}\left(1/ \tan^4 \beta \right),
\label{eq:mz}
\end{eqnarray}
where we have expanded in $1/\tan \beta$, so in the MSSM fine-tuning of the $Z$-boson mass arises principally from the parameters $\mu$ and $m_{H_u}$. Indeed, applying Eq.~(\ref{eq:mz}) to Eq.~(\ref{eq:FTdef}), the fine-tuning from $\mu$ alone is 
\begin{eqnarray}
\Delta_{\mu} \approx \frac{4 \lvert \mu \rvert^2}{M^2_Z},
\label{eq:FTmu}
\end{eqnarray}
which indicates that we need $\mu \lesssim \sqrt{5/2} M_Z \approx 150\,{\rm GeV}$ if we want to keep $\Delta_{\mu} \lesssim 10$.  Obviously $\sqrt{-m_{H_u}^2}$ must then also be small to give the correct $Z$-boson mass ($m_{H_u}^2$ is typically negative). However, in our $SU(5)$ GUT model, $m_{H_u}^2$ is not a free parameter, but is a polynomial function of the input parameters,
\begin{equation}
m^2_{H_{u}} = f\left( m_{\overline{5}}^{(3)}, m_{10}^{(3)} , K_{\overline{5}}, K_{10}, m_{\overline{5}^{\prime}}, m_{5^{\prime}}, M_3, \rho_1, \rho_2, a_{\overline{5}^{\prime}}, a_{5^{\prime}} \right),
\label{eq:mHu}
\end{equation}
with the  largest contributions arising from $m_{10}^{(3)}$, $m_{5^{\prime}}$, $M_3$ and $a_{5^{\prime}}$~\cite{Antusch:2012gv,Horton:2009ed}. If the dimensionful input parameters are ${\cal O} \left({\rm TeV} \right)$ or higher, motivated by the desire to avoid the LHC direct searches described in Sec.~\ref{sec:expc}, then small fluctuations in them will generally cause large fluctuations in our small $m_{H_u}^2$, which in turn spoils the $Z$-boson mass prediction and generates fine-tuning.

There are two potential ways out of this dilemma while still maintaining small $\mu$. Firstly one might imagine a scenario with ${\cal O} \left({\rm TeV} \right)$  dimensionful input parameters such that the contributions to the derivative in Eq.~(\ref{eq:FTdef}) just happen to cancel. The smallness of the $Z$-boson mass would be a coincidence, but one that was stable to local fluctuations. Unfortunately, as we shall see in Sec.~\ref{sec:ugm}, a scan over parameter space looking for such scenarios with universal gaugino masses found no examples with fine tuning less than $1000$. In Sec.~\ref{sec:nugm} we will see that we can do significantly better if we allow the gaugino masses to deviate from universality at the GUT scale, but fine-tuning is still sizable.

A second possibility would be if the dimensionful input masses were not ${\cal O} \left({\rm TeV} \right)$ at all, but actually rather small. Then their natural fluctuations would be small and the fluctuations of $m_{H_u}^2$ and thus fine-tuning would be reduced. In order to avoid the direct LHC searches one would have to generate sizable electroweak scale soft masses via the RGE evolution. Although this turns out to be rather easy to do for the scalar masses, it is unfortunately not possible for the gaugino masses. The leading order contribution to the gaugino RGE is proportional to the gaugino mass itself, so if the gaugino mass is small at high scales, it is always small. In contrast, the leading order sfermion RGEs contain the gaugino masses, which, if sufficiently large, can push the sfermion masses to TeV scales at low energies. So while one may reduce (or remove entirely) the fine-tuning arising from the scalars, one will still have fine-tuning from the gauginos.

To move forward, we will here take a constructive approach and regard fine-tuning as an indicator of new physical mechanisms. Since the fine-tuning in $\mu$ seems to be unavoidable, as discussed above, we will regard this as evidence that $\mu$ should not be regarded on the same footing as the soft supersymmetry breaking parameters in the theory. Indeed, the origin of $\mu$ is still one of the unsolved problems in supersymmetry; it is present in the superpotential before supersymmetry breaking, so {\it a priori} should know nothing about the electroweak scale. This is the well known $\mu$-problem, and suggests an effective $\mu$ parameter generated (possibly at high scales) by some unknown mechanism. The most famous example of such a mechanism is the Next-to-Minimal Supersymmetric Standard Model (NMSSM) (for a review, see~\cite{Miller:2003ay}) which introduces a new Higgs scalar field, $S$, that couples to the two MSSM Higgs doublets. This generates an effective $\mu$-term  when $S$ gains a vacuum expectation value, $\mu = \lambda \langle S \rangle$, where $\lambda$ is the coupling of the new scalar to the doublets. Alternatively $\mu$ may be generated by the F-term vacuum expectation value of a hidden sector field~\cite{Soni:1983rm}, $\mu = \langle F_X \rangle / M_{P}$, in a similar way to the soft supersymmetry breaking masses. However, neither of these suggestions would solve this fine-tuning problem: in the NMSSM, $\mu$ is proportional to $\lambda$ so one still has fine-tuning when varying $\lambda$; if $\mu$ is derived from an F-term one still has to fine-tune $\langle F_X \rangle$, its dimension $[{\rm mass}]^2$ only gaining us a factor of two due to the logarithmic form of Eq.~({\ref{eq:FTdef}).

Nevertheless, in this study, we will assume that some mechanism exists for generating an effective $\mu$ at the high scale that is insensitive to fluctuations in the true fundamental parameters and therefore does not provide a source of fine-tuning. Note that such a mechanism would not itself entirely solve the fine-tuning problem, since one must still require that the $m_{H_u}^2$, which contributes to $M_Z^2$ though Eq.~(\ref{eq:mz}), is also insensitive to variations in the fundamental GUT scale parameters. 

We will similarly consider that the ratios of the gaugino masses $\rho_1$ and $\rho_2$ must also have their origin in some underlying mechanism, otherwise, as we shall see in Sec.~\ref{sec:nugm},  they will also generate a large fine-tuning. Several mechanisms have been proposed in order to fix these ratios, and we have already discussed how these can be generated by non-trivial representations of hidden sector fields in Sec.~\ref{subsec:gugino}. Additionally,  orbifolds~\cite{Horton:2009ed, Brignole:1993dj} could be responsible for the non-universality of gaugino masses. We will explore both these possibilities in Sec.~\ref{sec:sfgmr}. Our fine-tuning is then only measured in  terms of the soft supersymmetry breaking parameters at the GUT scale. 

%%%%%%%%%%%%%%%%%
\section{Universal Gaugino Masses}
\label{sec:ugm}

We will first study scenarios with universal gaugino masses, $\rho_1 = \rho_2 = 1$. We randomly chose our input parameters within the ranges given in Sec.~\ref{sec:scans} and run them down to the electroweak scale using the full two-loop RGEs within SOFTSUSY. We set the electroweak scale to be \mbox{$M_z = 91.1876\,{\rm GeV}$} and the top quark pole mass to be $m_t = 173.4$ GeV. We do not force {\it exact} gauge couplings unification in order to allow possible percent level shifts due to threshold corrections at unification scale, as well as shifts arising from possible higher dimensional operators. 

As a preliminary cut, to avoid unnecessary computation, we discard scenarios with Higgs boson masses outside the range $122.6-127\,{\rm GeV}$, and also any scenarios that do not respect the LHC direct and XENON100 (2$\sigma$) bounds as described in Sec.~\ref{sec:expc}. 
We ensure that our scenarios have a stable vacuum using the conditions proposed by Casas, Lleyda and Mu\~noz in \cite{Casas:1995pd}. Specifically, we implement the unbounded from below (UFB) constraints UFB-1,2,3 and the charge and colour breaking (CCB) minima constraint CCB-1. In the interest of computational efficiency we take a simplified approach to the CCB-2,3 constraints and implement the simple cuts $\left| a_{\overline{5}'}/m_{\overline{5}} \right| \lesssim 3$, $\left| a_{5'}/m_{10} \right| \lesssim 3$ and $\left| a_{\bar 5'}/m_{10} \right| \lesssim 3$ to ensure they are satisfied. At this stage we also discard scenarios with a charged LSP (the majority of these have a stau LSP, caused by a low value of $m_{10}$). Out of 2,000,000 initial attempts, this leaves approximately 57,000 scenarios in our scan. 

We then use the electroweak scale outputs of SOFTSUSY as inputs for micrOMEGAS to generate predictions for the remaining experimental observables, such as the relic density, and derive a value of ${ P}_{\rm tot}$ for each scenario. Requiring ${ P}_{\rm tot}>10^{-3}$ reduces the number of viable scenarios to 306, the vast majority of which have a Dark Matter relic density below the constraint described in Sec.~\ref{sec:expc}; only 30 scenarios have the preferred relic density. 

Fig.~\ref{fig:Umutanb} shows the distribution of these surviving points in $\mu$ and $\tan \beta$, where scenarios with a Dark Matter relic density below the 2$\sigma$ relic density bounds are shown in blue, while those with the preferred value are shown in green. Most scenarios are in the region \mbox{$150 \gtrsim \mu \gtrsim 600\,{\rm GeV}$}, where  the dark matter candidate is mainly a neutralino dominated by its Higgsino component with mass $m_{\tilde{\chi}^0_1} \approx \mu$. These scenarios generally have large values of $m_{5^{\prime}} \gtrsim 2\,{\rm TeV}$, which force a low value of $m^2_{H_u}$ due to the RGE running, and in a turn a relatively low value of $\mu$ from the $Z$-boson mass constraint. 30 scenarios have the preferred relic density: 28 of these have bino dominated neutralinos as the LSP; only 2 have higgsino dominated neutralinos as the LSP (the two green points in the figure with smallest $\mu$).
\begin{figure}[ht!]
\centering
\includegraphics[width=0.5\textwidth]{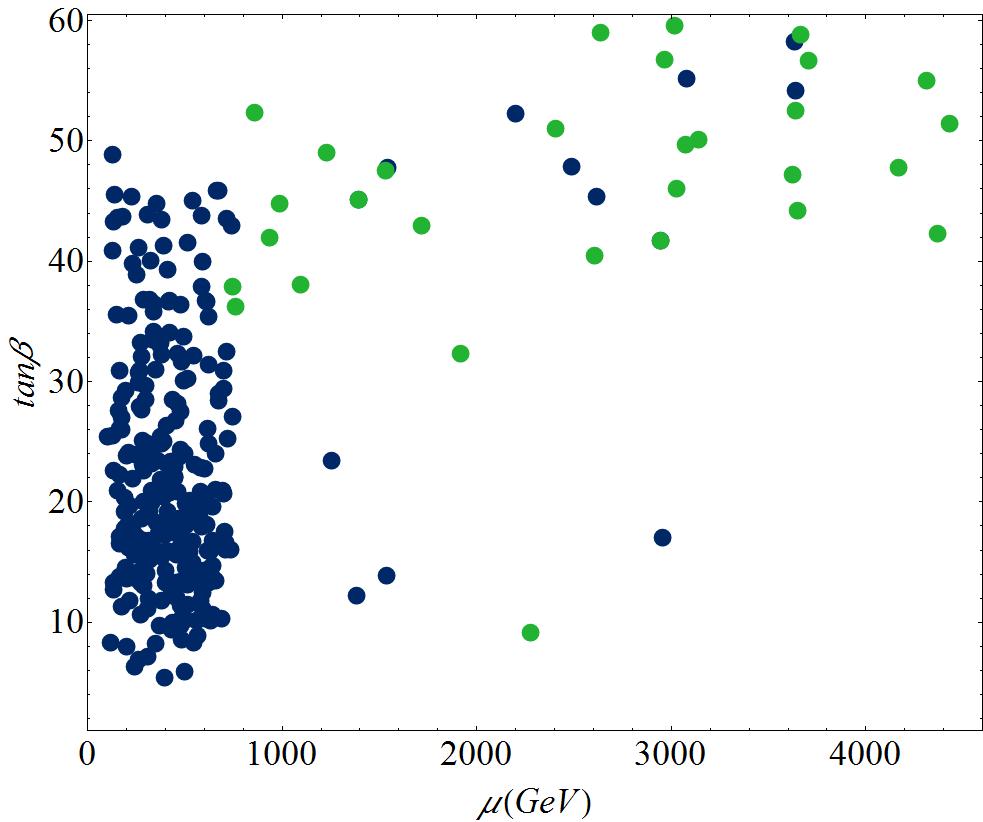} 
\caption{\it Viable universal gaugino mass scenarios in the $\mu$ - $\tan \beta$ plane. Blue points represent scenarios with a Dark Matter relic density below $2\sigma$ bounds, while green points have the preferred relic density. }
\label{fig:Umutanb}
\end{figure}   

In Fig.~\ref{fig:Ustophiggs} we also show the viable scenarios with respect to the physical stop masses, and the Higgs boson and its pseudo-scalar partner. The lightest stop $\tilde{t}_1$ we found was $461\,{\rm GeV}$ (this is the blue point furthest to the left) though this has a Dark Matter relic density below observations. The lightest stop with the preferred relic density has mass $534\,{\rm GeV}$ (the furthest left green point). The other characteristics of these two scenarios can be found in the benchmarks $\rm{BP1}SU(5)_{\mathbf{1}}$ and $\rm{BP2}SU(5)_{\mathbf{1}}$ described in Sec.~\ref{sec:BP}. From  Fig.~\ref{fig:Ustophiggs} (right) we see that we can produce a sufficiently heavy Higgs boson, but we require a CP-odd Higgs mass, $m_{A}$ in the approximate region of $1$--$4.5\,{\rm TeV}$. In a recent work by Baer et al \cite{Baer:2012cf} acceptable solutions were found with $m_{A}$ in the interval $150$--$1500\,{\rm GeV}$, where we find very few viable solutions. However, Bear et al consider $m_A$ as an input and restrict to this range to generate a scan over parameter space; by contrast our $m_A$ is an output derived from the running of the GUT scale parameters. It is possible that with our much wider parameter scan we fail to find viable solutions with $m_A < 1\,{\rm TeV}$, but would find them if we greatly increased our initial number of scenarios tested.
\begin{figure}[ht!]
\centering
\includegraphics[width=0.48\textwidth]{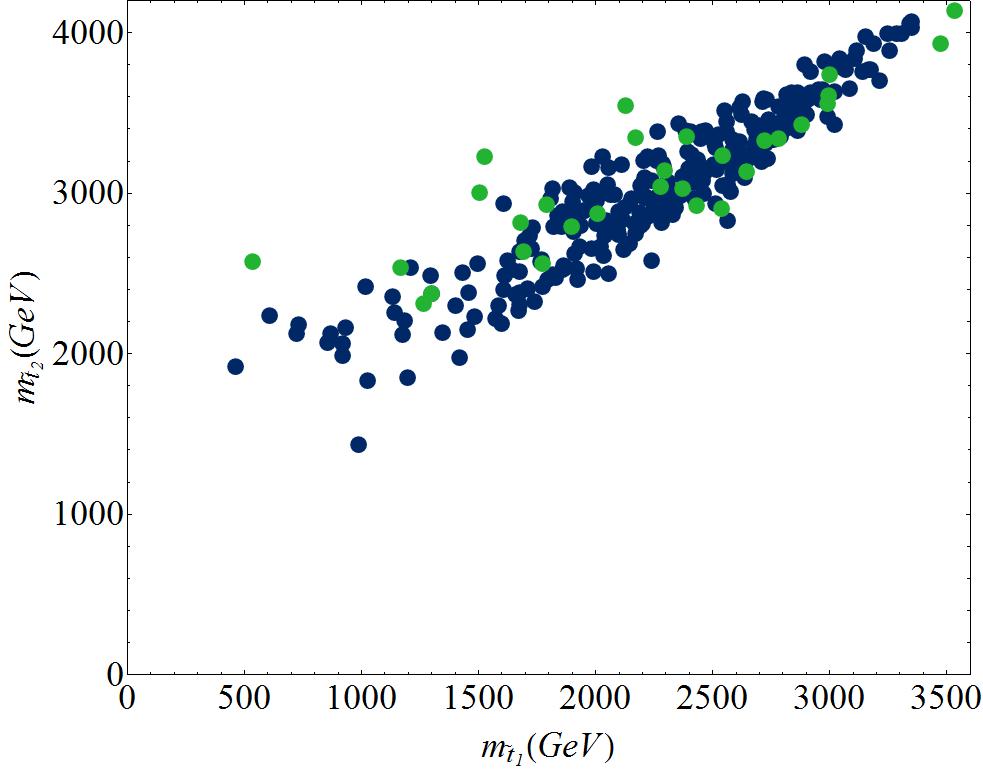}
\includegraphics[width=0.48\textwidth]{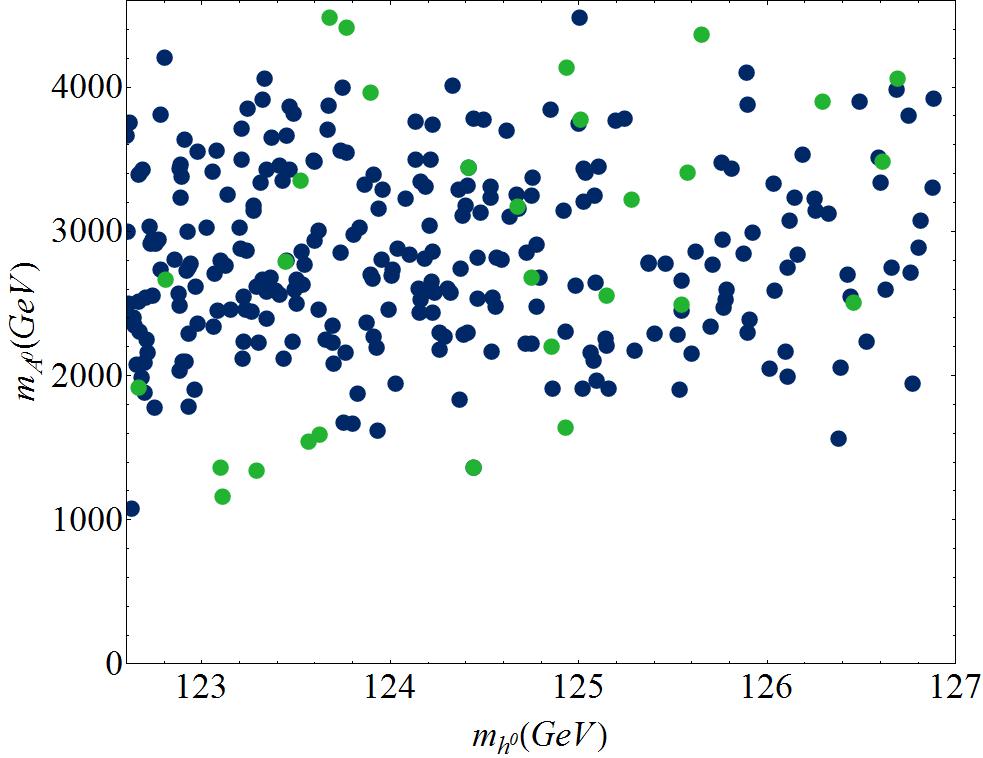}
\caption{\it Viable universal gaugino mass scenarios in the stop mass (left) and the lightest scalar - pseudoscalar mass (right) planes, with colours as in Fig.~\ref{fig:Umutanb}.}
\label{fig:Ustophiggs}
\end{figure}   

\begin{figure}[ht!]
\centering
\includegraphics[width=0.48\textwidth]{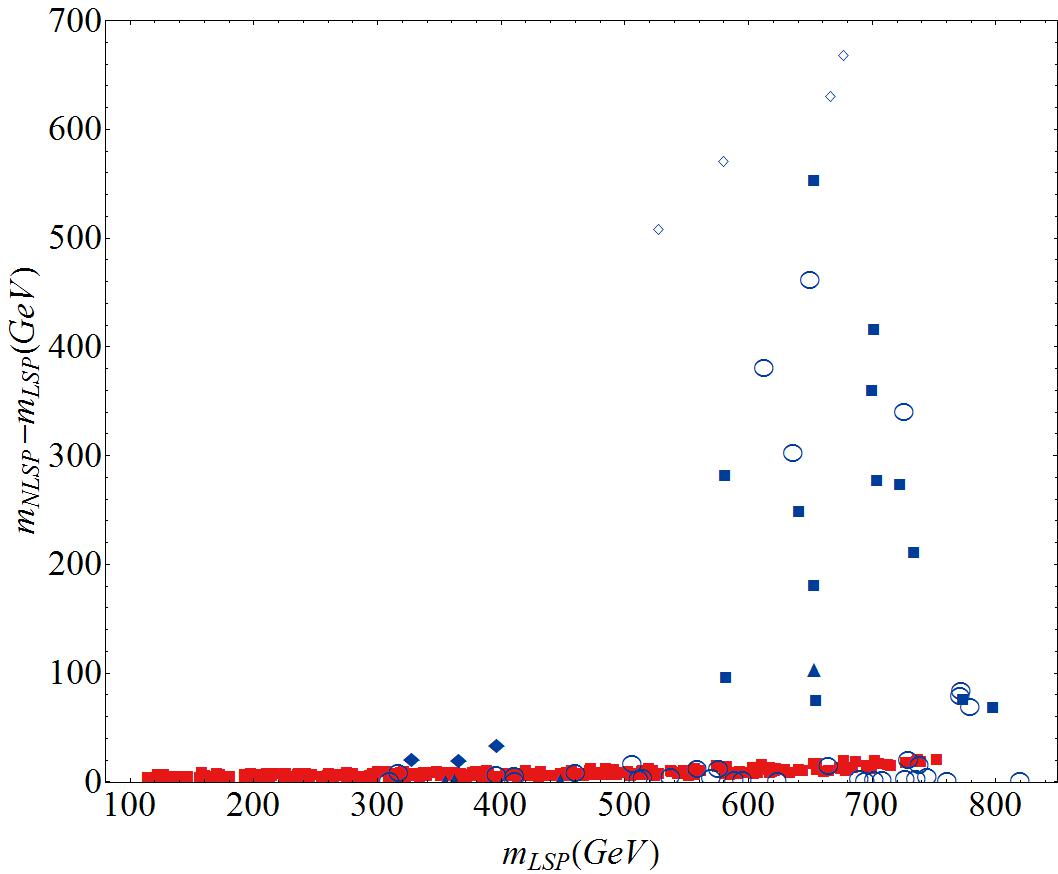}
\includegraphics[width=0.47\textwidth]{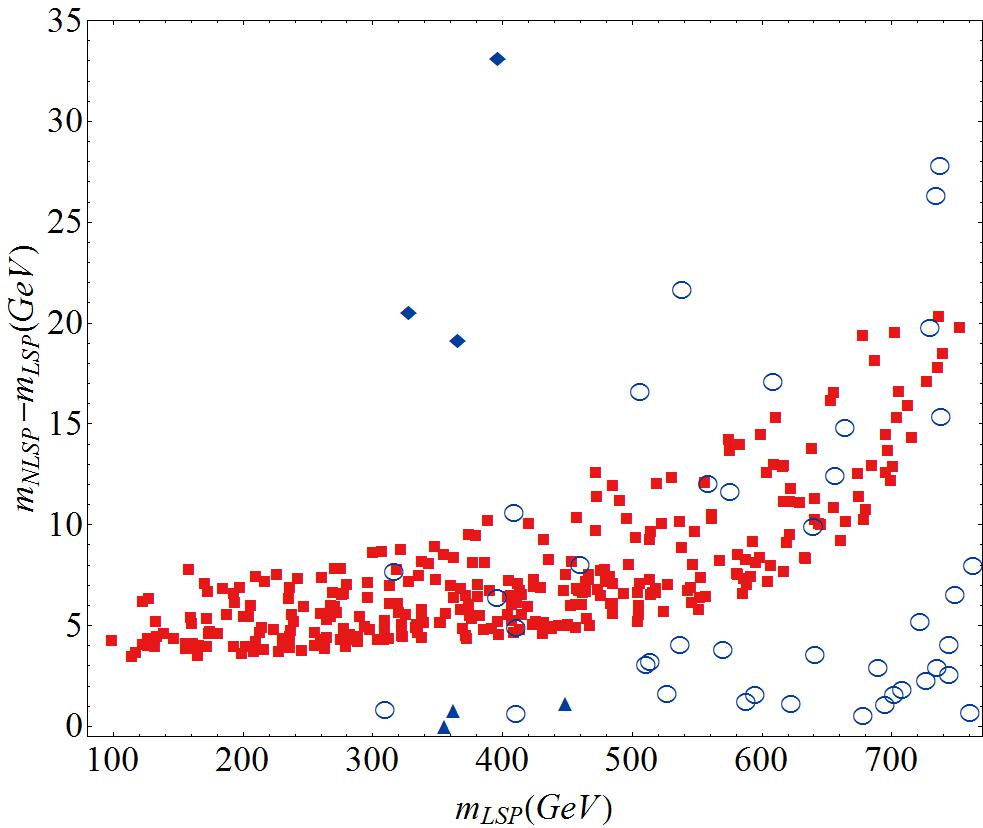}
\label{fig:ULSPtype}
\caption{\it Solutions in the plane of LSP mass vs.~the NLSP-LSP mass splitting for universal gaugino mass scenarios. The colour indicates the flavour of LSP, with red and blue denoting higgsino and bino dominated Dark Matter respectively. The shape indicates the flavour of NLSP; squares, diamonds, triangles and circles denote chargino, stop, sneutrino and stau NLSP respectively. The right-hand plot is a zoomed in version of the left-hand plot.}
\label{fig:UDM}
\end{figure}
The LHC constraint on the gluino mass of $M_{\tilde{g}} \gtrsim 800\,{\rm GeV}$ imposes a lower bound of about $M_{1/2} \gtrsim 300\,{\rm GeV}$ on the common gaugino mass at the high scale. $M_{1/2} \approx 300\,{\rm GeV}$ would result in a bino dominated neutralino with mass around $150\,{\rm GeV}$. If the LSP, this would give too high a Dark Matter relic density unless one has an approximately degenerate Next-to-Lightest Supersymmetric Particle (NLSP) to facilitate coannihilation, or an appropriate particle at twice the LSP mass to provide resonant decay. Unfortunately we find no such scenarios that evade the experimental constraints and instead find that scenarios with a gluino near the LHC bound require a higgsino dominated neutralino as Dark Matter with a chargino as NLSP. These are the red squares shown in the low mass region of Fig.~\ref{fig:UDM}.
When $M_{1/2}$ is raised to $700\,{\rm GeV}$ or greater, the bino mass becomes greater than about $300\,{\rm GeV}$} and then we do indeed find viable scenarios with a bino dominated LSP and an acceptable Dark Matter relic density. All our viable scenarios are shown in Fig.~\ref{fig:UDM}. \\

%%%%%%%%%%%%%%%%%

We have so far seen that for $SU(5)$-inspired models with universal gaugino mass one has plenty of solutions that survive the experimental constraints and vacuum stability conditions, including an acceptable relic density of Dark Matter. Now we will examine these scenarios to see if they have significant fine-tuning from sources other than $\mu$. 

\begin{figure}[ht!]
\centering
\includegraphics[width=0.48\textwidth]{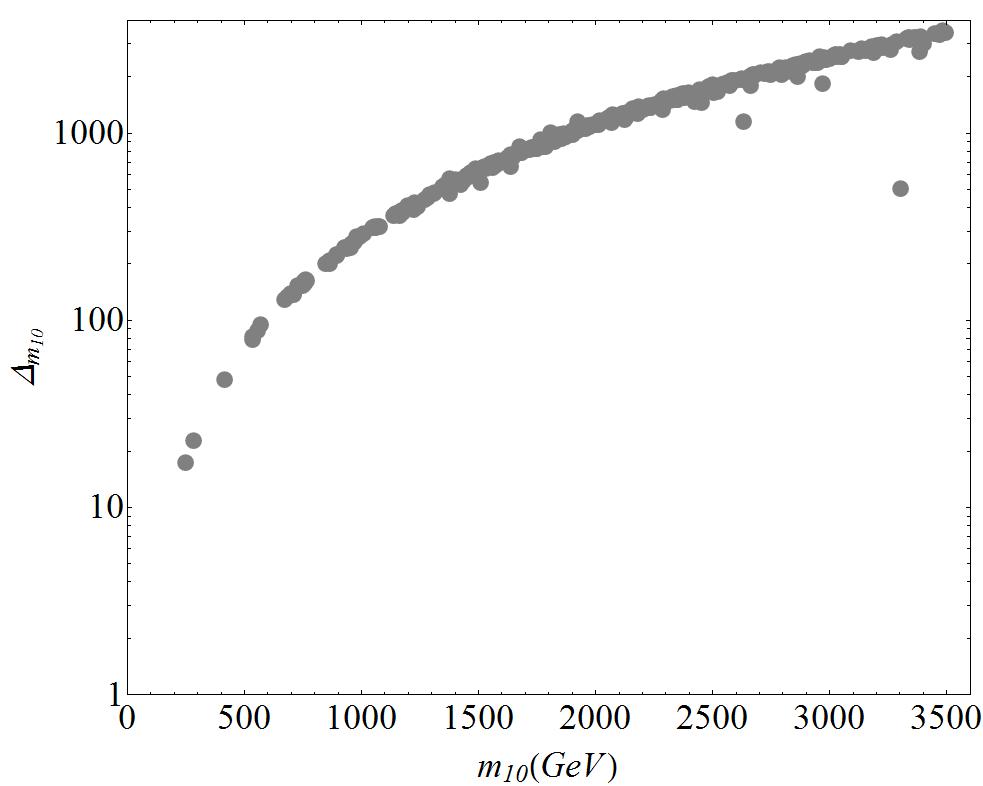}
\includegraphics[width=0.48\textwidth]{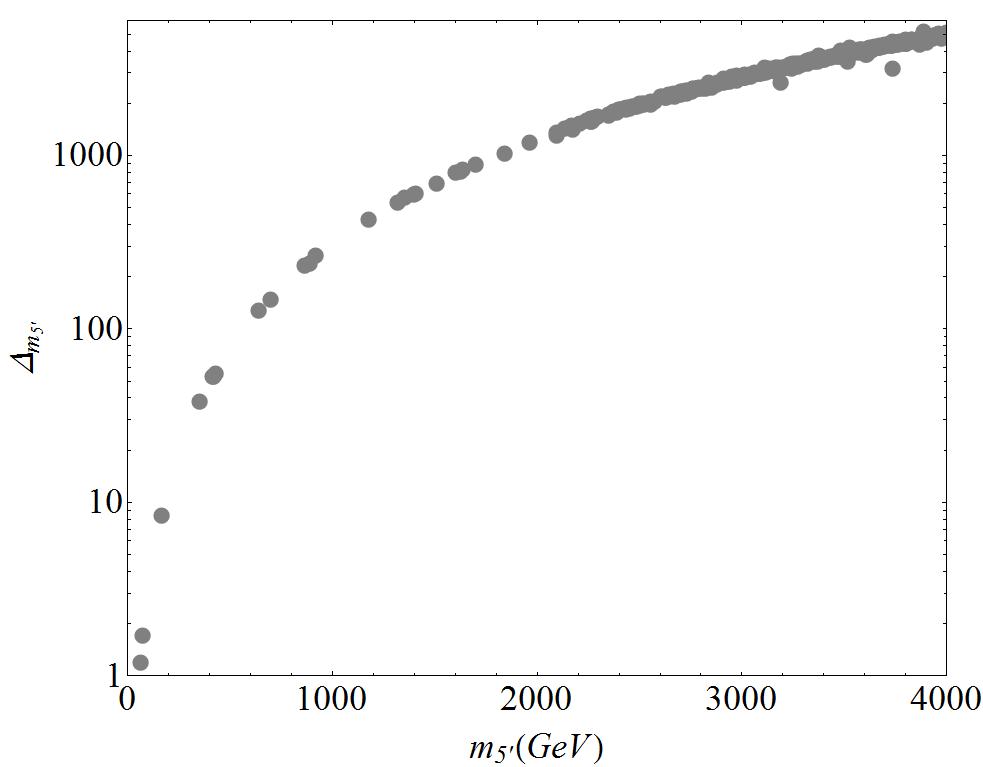}\\[4mm]
\includegraphics[width=0.48\textwidth]{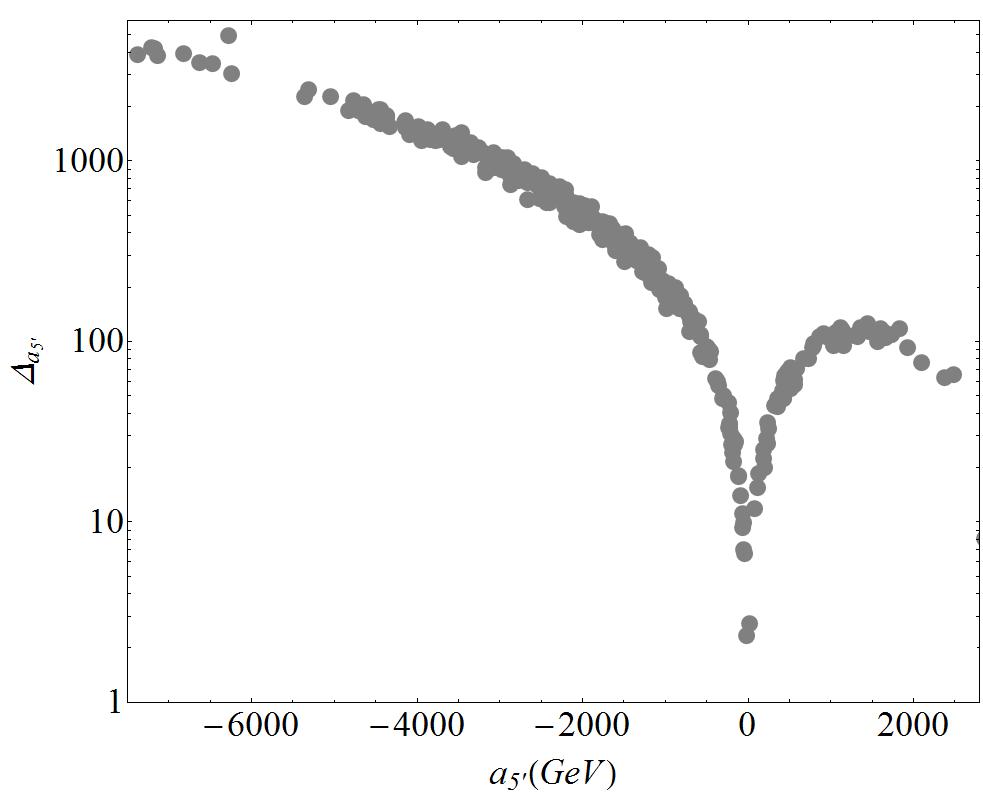}
\includegraphics[width=0.48\textwidth]{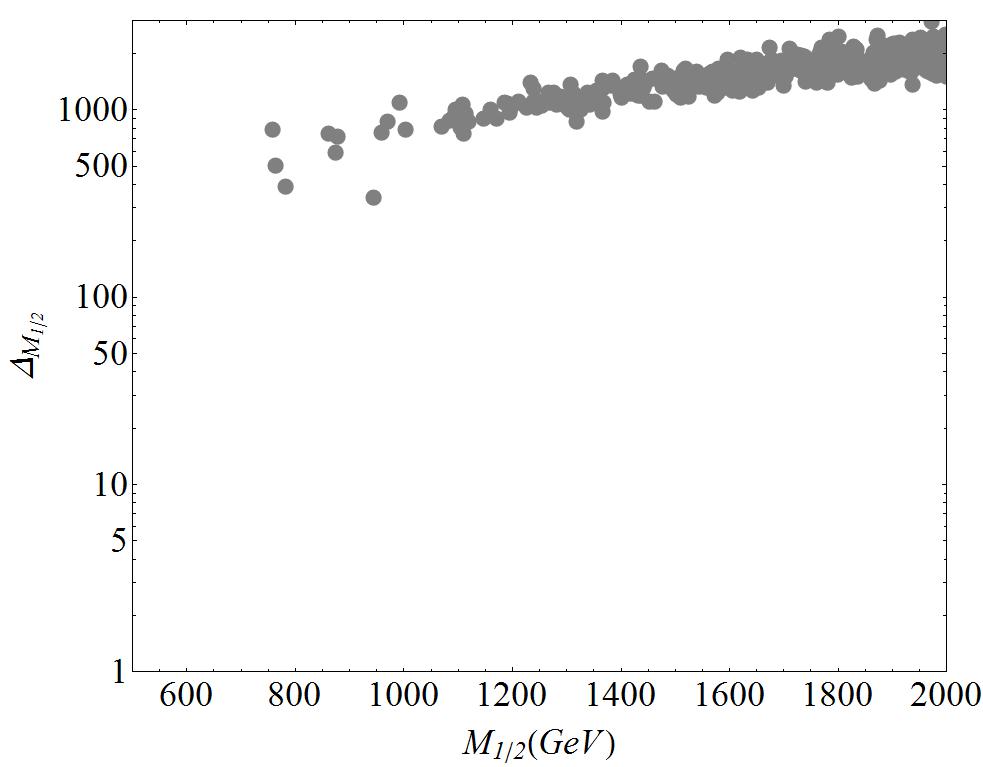}
\caption{\it Fine-tuning in $M_Z$ with respect to the input parameters $m_{10}$, $m_{5^{\prime}}$, $M_{1/2}$ and $a_{5^{\prime}}$ for universal gaugino mass scenarios.}
\label{fig:UDeltas}
\end{figure}

\begin{figure}[t!]
\centering
\includegraphics[width=0.5\textwidth]{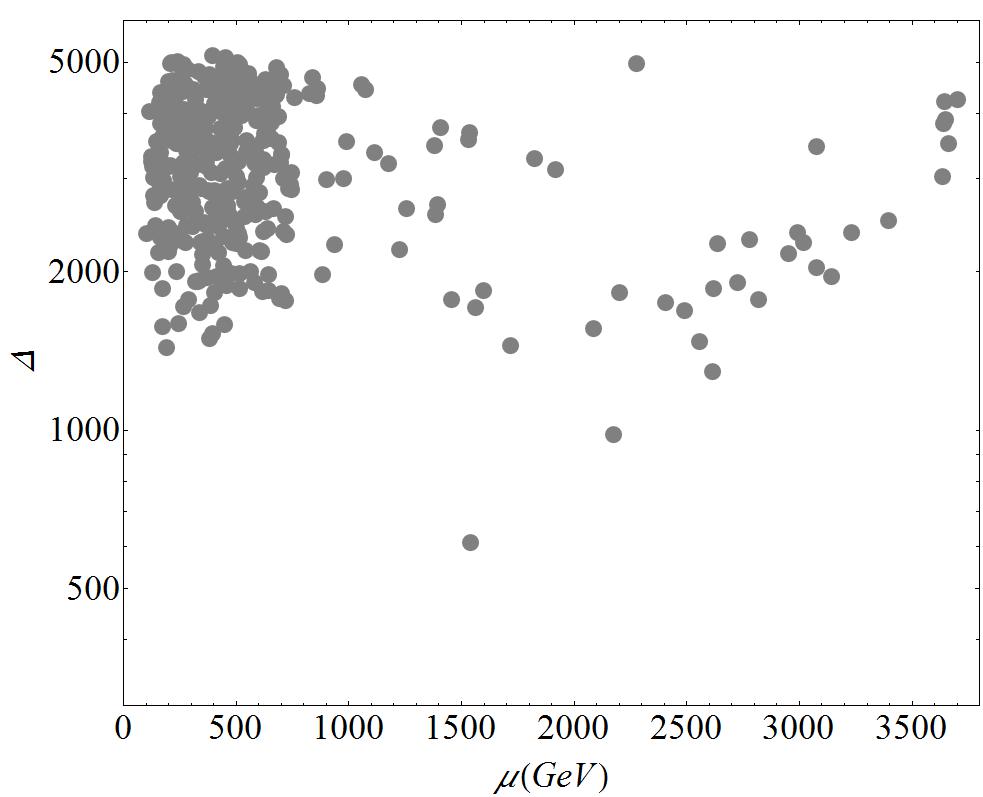}
\caption{\it The fine-tuning $\Delta$ compared to $\mu$ for universal gaugino mass scenarios.}
\label{fig:Umuft}
\end{figure}

In particular we focus on fine-tuning of $M_Z$ due to shifts in $m_{10}$, $m_{5^{\prime}}$, $M_{1/2}$ and $a_{5^{\prime}}$, which provide the dominant contribution to $m_{H_u}^2$. We use SOFTSUSY's implementation of fine-tuning throughout, which uses a discretised version of the definition in Eq.~(\ref{eq:FTdef}). The independent fine-tunings in these parameters are shown in Fig.~\ref{fig:UDeltas}. We see that the individual fine-tunings $\Delta_{m_{10}}$, $\Delta_{m_{5^{\prime}}}$ and $\Delta_{a_{5^{\prime}}}$ become small as their corresponding parameters are taken to zero, but we find no scenario with
$\Delta_{M_{1/2}}$ less than about $330$. 
This fine-tuning problem is exacerbated when these individual fine-tuning are combined into $\Delta$, which is defined as the maximum value of the four tunings for each scenario (recall we are discounting the fine-tuning with respect to $\mu$). In Fig.~\ref{fig:Umuft} we show this total fine-tuning in comparison to $\mu$, and see that for the majority of scenarios we never have $\Delta$ less than about 1300. The minimum value of $\Delta$ found was $611$ with a rather large value of $\mu$ (and thus $\Delta_\mu$). For viable scenarios in the region with Higgsino dominated dark matter, $100\, {\rm GeV} \lesssim \mu \lesssim 800\,{\rm GeV}$, $\Delta_{\mu}$ may have been tolerable but unfortunately the fine-tuning in the other parameters make these  unattractive. 

The results obtained in this section show that it is possible to obtain physically viable solutions for GUT scale $SU(5)$-inspired scenarios with universal gaugino masses. However, all the scenarios found have a significant degree of fine tuning. We do however note that one may be able to find additional solutions with low fine tuning with a more intensive search~\cite{Baer:2012cf}, though such scenarios are undoubtedly rare. 

%%%%%%%%%%%%%%%%%
\section{Non-Universal Gaugino Masses}
\label{sec:nugm}

We expand our analysis by allowing the gaugino masses at the GUT scale to depend on the (SM) gauge group. This requires the introduction of two extra parameters, $\rho_1$ and $\rho_2$ which  we vary in the interval $[-10, 10]$. We will continue to use the notation $M_{1/2}$ for the value of $M_3$ at the GUT scale in order to distinguish it from its value at other energies. 

\subsection{An Inclusive Scan}

We begin our study of non-universal gaugino masses with an inclusive scan over parameter space to seek regions of interest,  following a similar procedure to the universal gaugino mass scenarios described in Sec.~\ref{sec:ugm}. We increase the number of initial tries to 2,500,000 since we now have a larger parameter space to scan. After the preliminary Higgs mass cut, imposing the LHC and XENON100 direct search bounds, applying stability constraints and removing charged LSP scenarios we find only 22,418 scenarios (0.9\%) survive, in comparison to approximately 57,000 (3\%) for universal gaugino masses. This reduction in the number of accepted scenarios is due to the additional removal of scenarios with coloured dark matter in regions where \mbox{$M_{3} \ll M_{1,2}$}. However, we find that the surviving scenarios are more accommodating to both the additional experimental constraints and the relic abundance of Dark Matter. After requiring ${ P}_{\rm tot}>10^{-3}$ we find approximately 13,191 scenarios remain, 1581 of which have the preferred relic abundance of Dark Matter.

\begin{figure}[t!]
\centering
\includegraphics[width=0.45\textwidth]{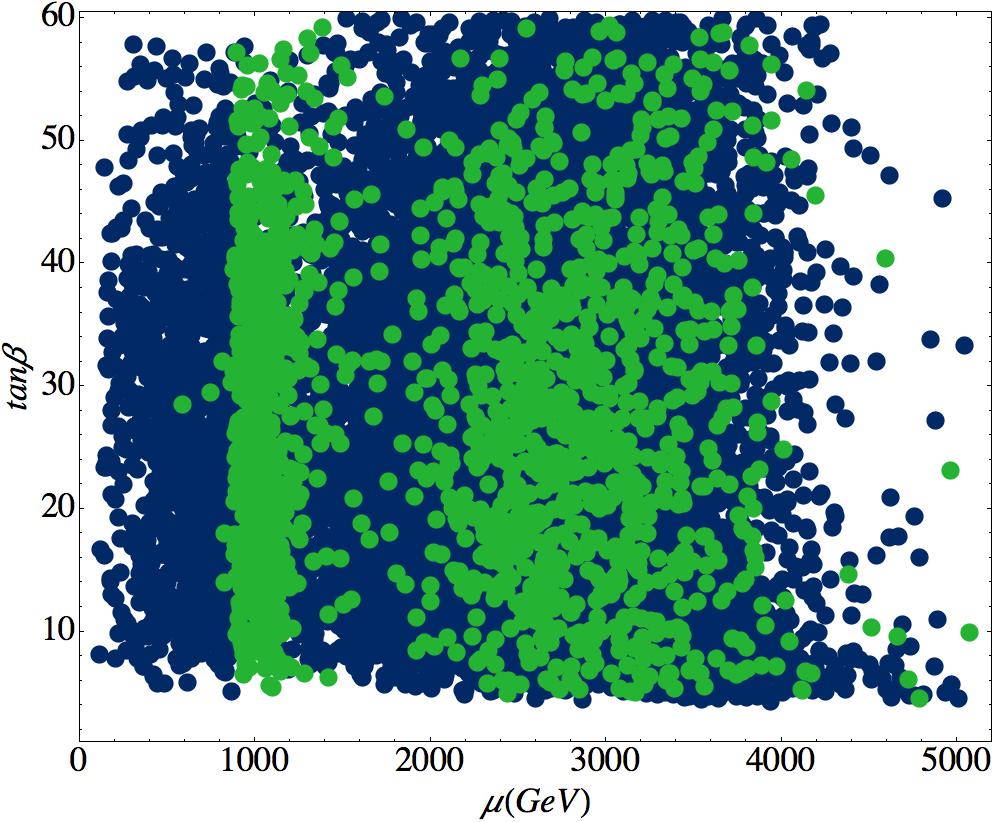}
\caption{\it Viable non-universal gaugino mass scenarios in the $\mu$-$\tan \beta$ plane, with colours as in Fig.~\ref{fig:Umutanb}.}
\label{fig:NUmutanb}
\end{figure}

The gaugino masses feed into the RGEs of all superpartners playing an important role on their evolution, so it is not surprising that the range of physical masses is extended by relaxing the universality constraint. We show the viable scenarios projected onto the $\mu$-$\tan \beta$ plane in Fig.~\ref{fig:NUmutanb}.  In contrast to the universal gaugino mass scenarios, we now have many examples of the preferred Dark Matter relic density, where the green band around $1\,{\rm TeV}$ predominantly represents scenarios with higgsino dominated Dark Matter. We find viable scenarios with stop masses ranging from few hundred GeV up to $6\,{\rm TeV}$, and a pseudoscalar Higgs mass extended to the interval $1.2-6\,{\rm TeV}$.

\begin{figure}[ht!]
\centering
\includegraphics[width=0.45\textwidth]{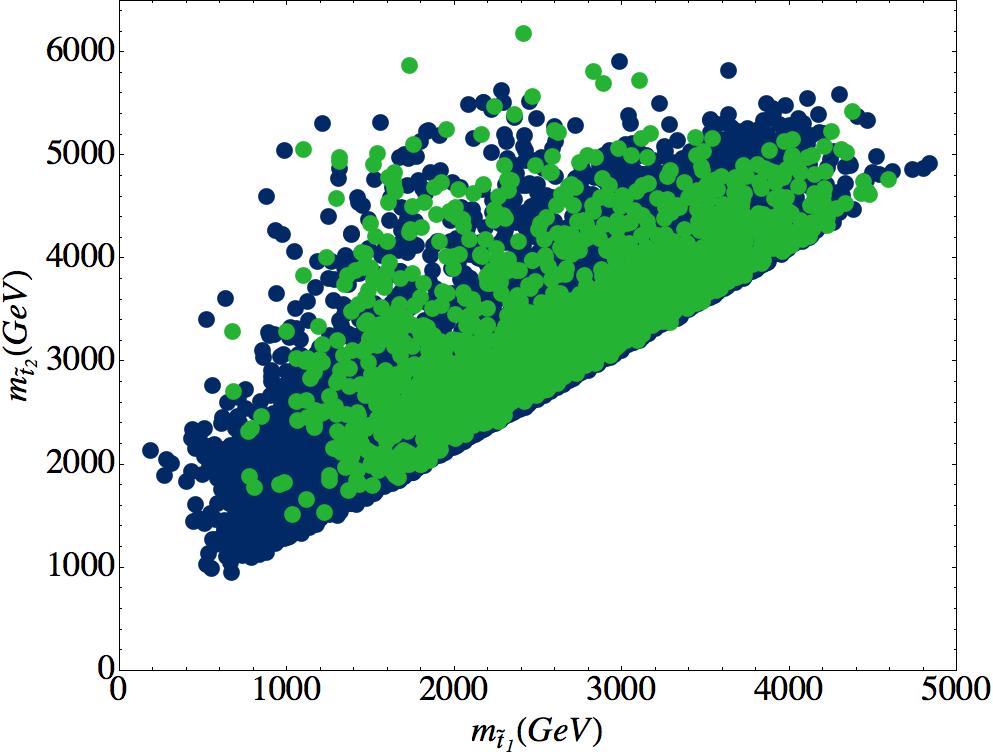}
\includegraphics[width=0.45\textwidth]{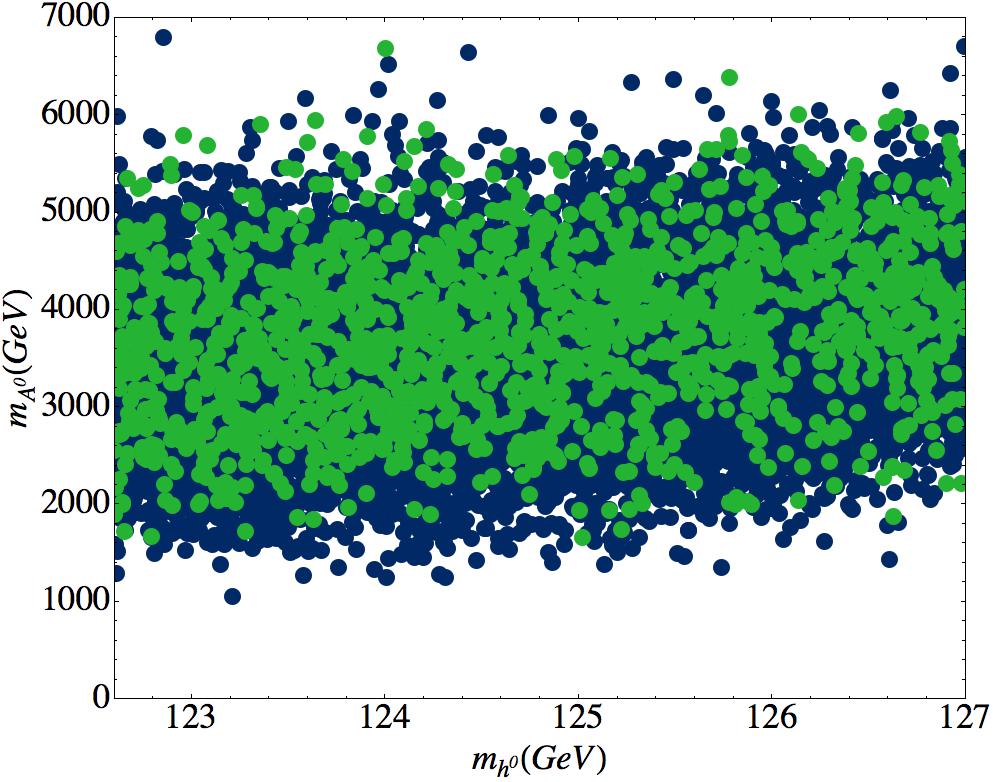}
\caption{\it Viable non-universal gaugino mass scenarios in the stop mass  (left) and the lightest scalar -- pseudoscalar mass  (right) planes, with colours as in Fig.~\ref{fig:Umutanb}.}
\label{fig:NUstophiggs}
\end{figure} 

\begin{figure}[h!]
\centering
\includegraphics[width=0.5\textwidth]{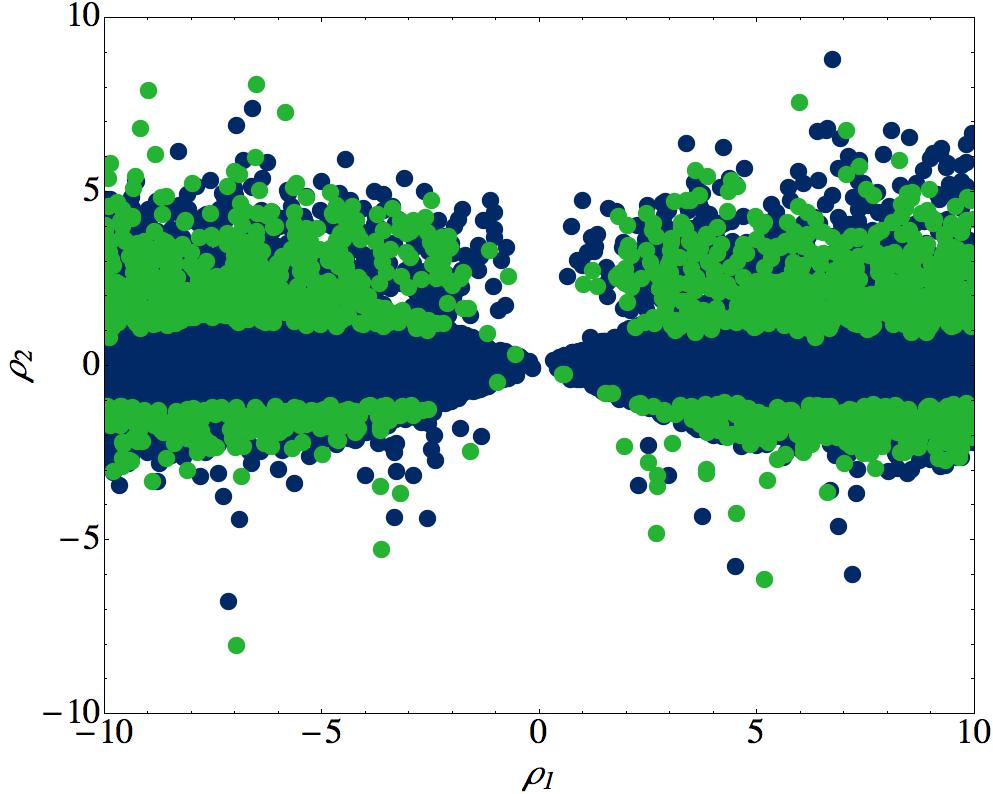}
\caption{\it Viable non-universal gaugino mass scenarios in the non-universality parameters $\rho_{1,2}$, with colours as in Fig.~\ref{fig:Umutanb}.}
\label{fig:NUrho}
\end{figure}

The values of stop masses and the scalar/pseudoscalar Higgs masses are shown in Fig.~\ref{fig:NUstophiggs}.
The scenarios with light sfermions (staus as well as stops) would be visible at the $14\,{\rm TeV}$ LHC. However, these solutions tend to have too little Dark Matter and we find very few scenarios with the preferred Dark Matter relic density while maintaining stops below $1\,{\rm TeV}$. 
The values of the non-universality parameters $\rho_{1,2}$ for viable scenarios are shown in Fig.~\ref{fig:NUrho}. Notice that there are very few viable scenarios around $\rho_1=\rho_2=1$ corresponding to universal gaugino masses. 

\begin{figure}[ht!]
\centering
\includegraphics[width=0.48\textwidth]{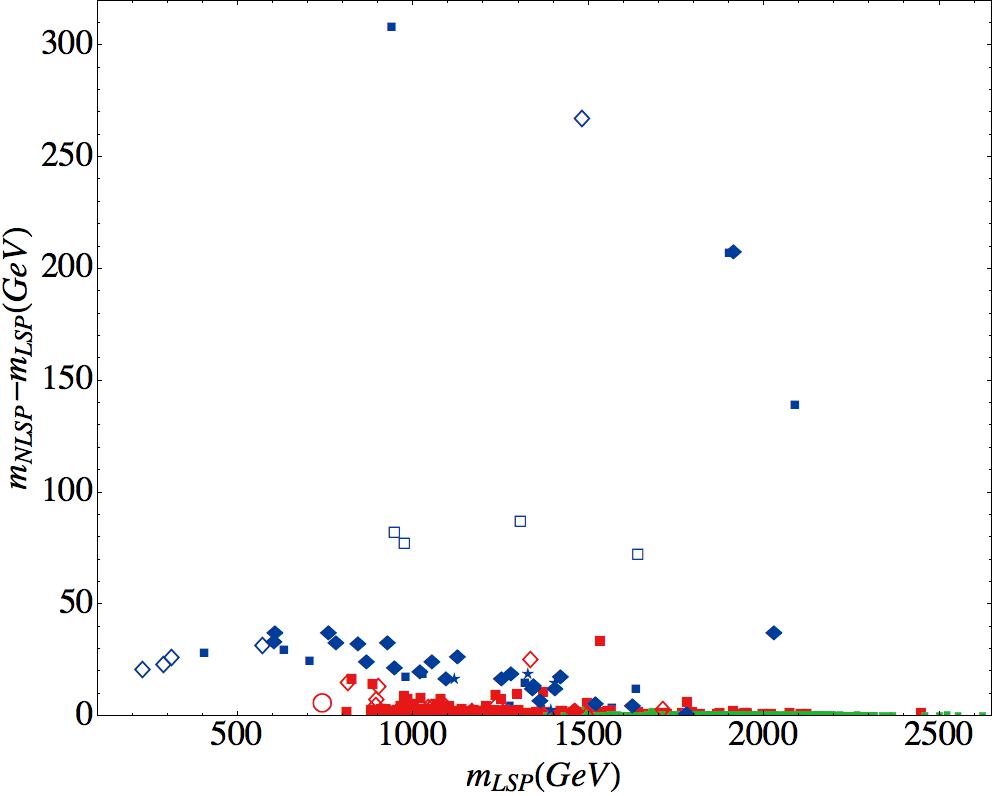}
\includegraphics[width=0.48\textwidth]{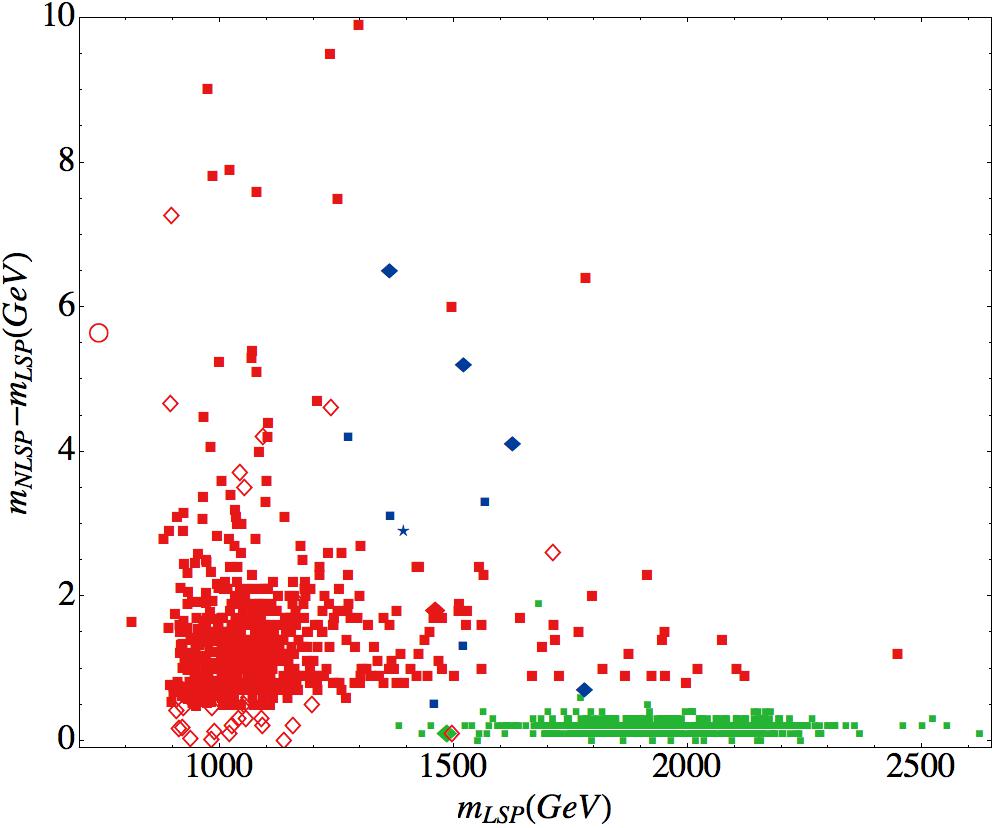}
\caption{\it Solutions in the plane of LSP mass vs.~the NLSP-LSP mass splitting for non-universal gaugino mass scenarios. The colour indicates the flavour of LSP, with red, blue and green denoting higgsino, bino and wino dominated Dark Matter respectively. The shape indicates the flavour of NLSP; filled squares, empty squares, filled diamonds, empty diamonds, circles and stars denote chargino, gluino, stop, neutralino, stau and sbottom NLSP respectively. In contrast to Fig.~\ref{fig:UDM}, to keep the figure becoming too densely populated, we only show scenarios with the preferred Dark Matter relic density. The right-hand plot is a zoomed in version of the left-hand plot.}
\label{fig:NUDM}
\end{figure}

In Fig.~\ref{fig:NUDM} we show the identity and masses of the LSP and NLSP for scenarios with the preferred relic density and now see many extra possibilities for LSP-NSLP pairings. Indeed the non-universality of gaugino masses now allows $M_2$ to be smaller than $M_1$, so we may also have wino dominated Dark Matter, and this can provide the correct relic density for higher LSP masses. As for the universal gaugino mass scenarios, the LSP and NLSP are typically close in mass in order to encourage co-annihilation but for bino dominated Dark Matter it is possible to have the NLSP as much as $300\,{\rm GeV}$ heavier than its LSP. (This particular scenario has a heavy Higgs boson twice the LSP mass allowing Dark Matter annihilation via a Higgs resonance.) \\

\begin{figure}[t!]
\centering
\includegraphics[width=0.6\textwidth]{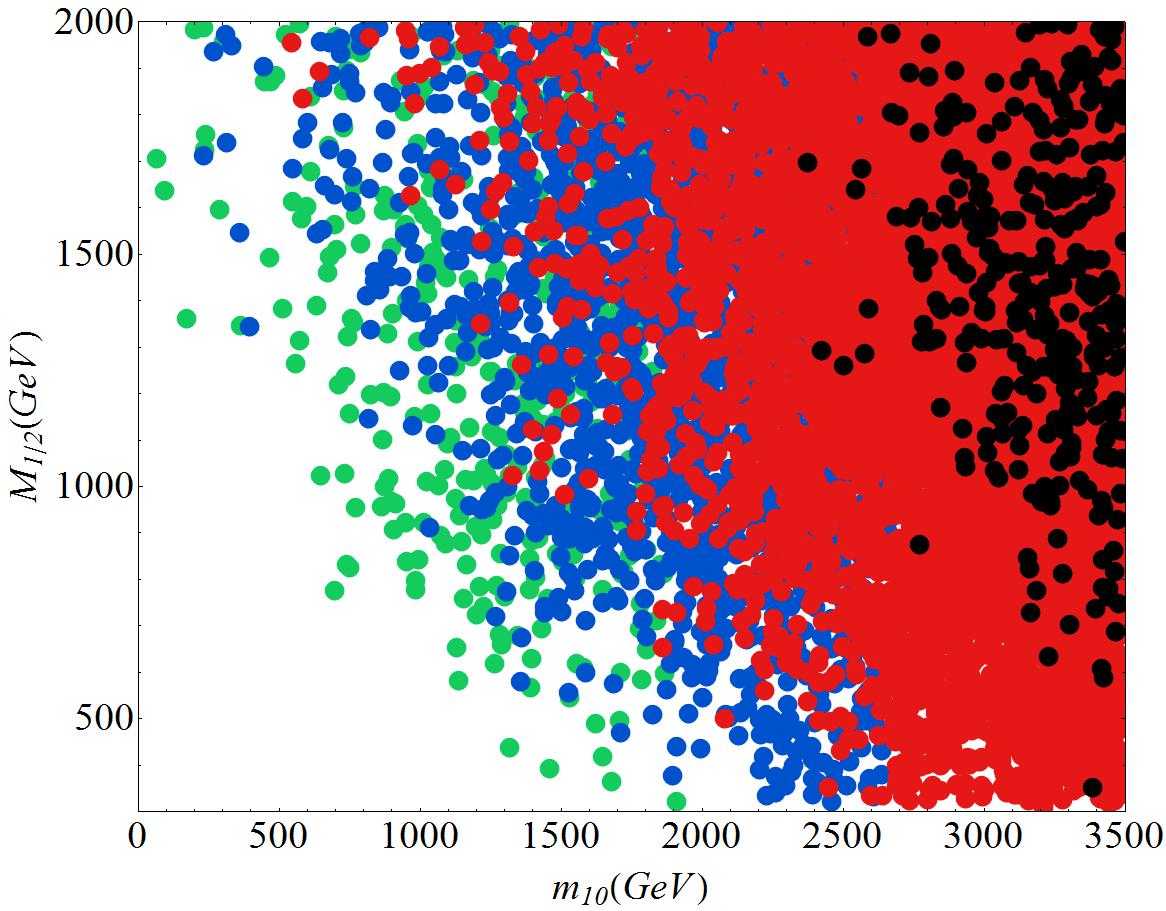}
\caption{\it Fine-tuning as a function of the input masses $m_{10}$ and $M_{1/2}$ for non-universal gaugino mass scenarios.
Green points represent scenarios with $\Delta \leq 1000$; blue points $1000 < \Delta \leq 2000$; red points $2000 < \Delta \leq 5000$; and black points $\Delta > 5000$.}
\label{fig:NUm10m12Delta}
\end{figure}

Although fine-tuning can be greatly reduced when the gaugino mass constraints are relaxed, there is still significant fine-tuning for much of the parameter space, and we find only one point with  $\Delta < 100$. This has $\mu \approx 500\,{\rm GeV}$ and a fine-tuning of approximately $60$. Recall that the fine-tuning of $\mu$ is not included in $\Delta$; the fine-tuning in $\mu$ as given by 
 Eq.~(\ref{eq:FTmu}) is of order $120$. 
In Fig.~\ref{fig:NUm10m12Delta} we show the fine-tuning in the $m_{10}$-$M_{1/2}$ plane. The white area to the bottom-left of this plot is excluded by the experimental constraints. We see that increasing $m_{10}$ very quickly gives unpalatable values for the fine-tuning, but increasing $M_{1/2}$ is not so problematic. This leads us to speculate that  low values of the (GUT scale) soft scalar masses may provide attractive scenarios as long as a large $M_{1/2}$ feeds their evolution, making the scalars heavy enough to avoid the LHC constraints.

\begin{figure}[h!]
\centering
\includegraphics[width=0.44\textwidth]{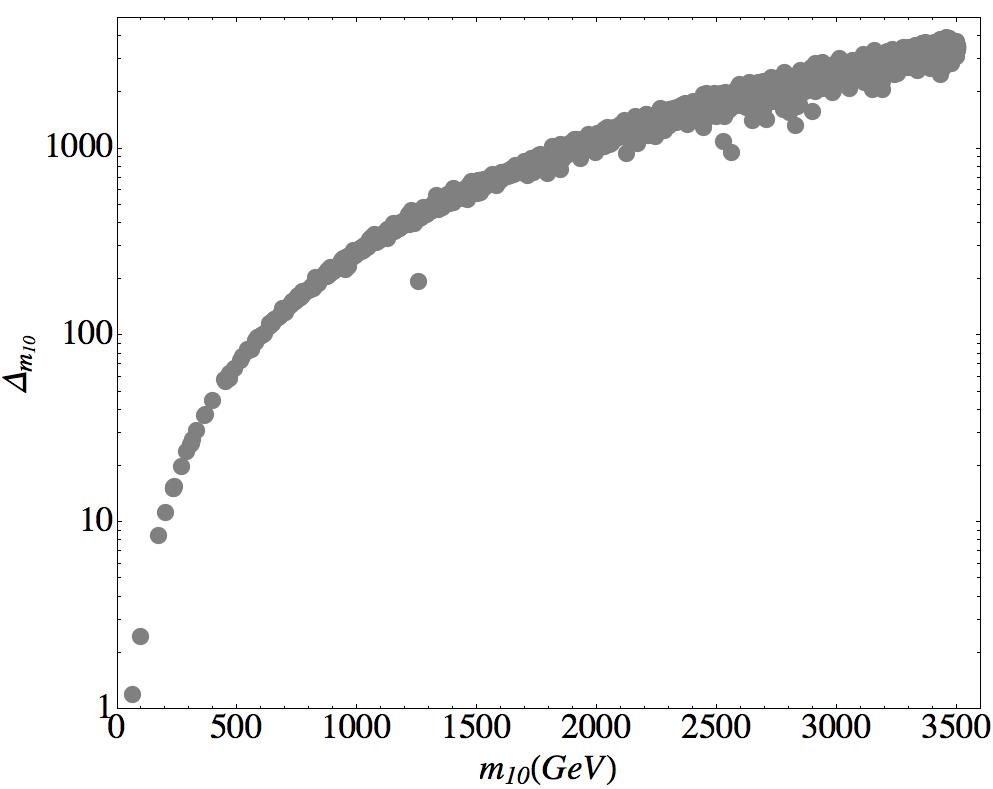}
\includegraphics[width=0.44\textwidth]{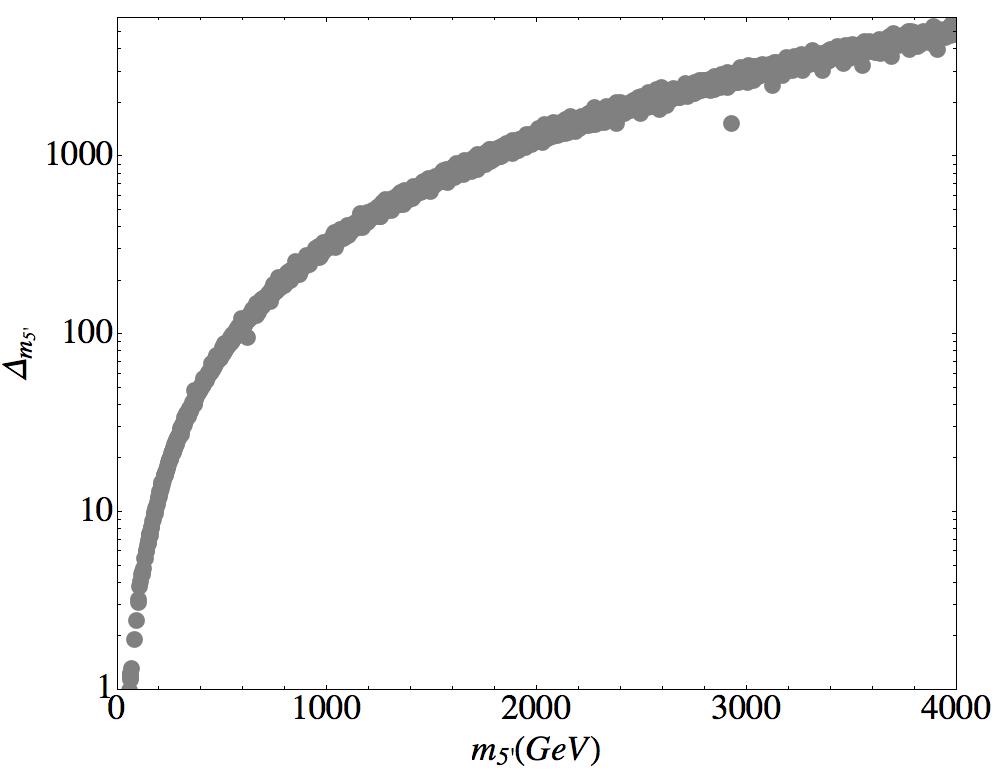}
\includegraphics[width=0.44\textwidth]{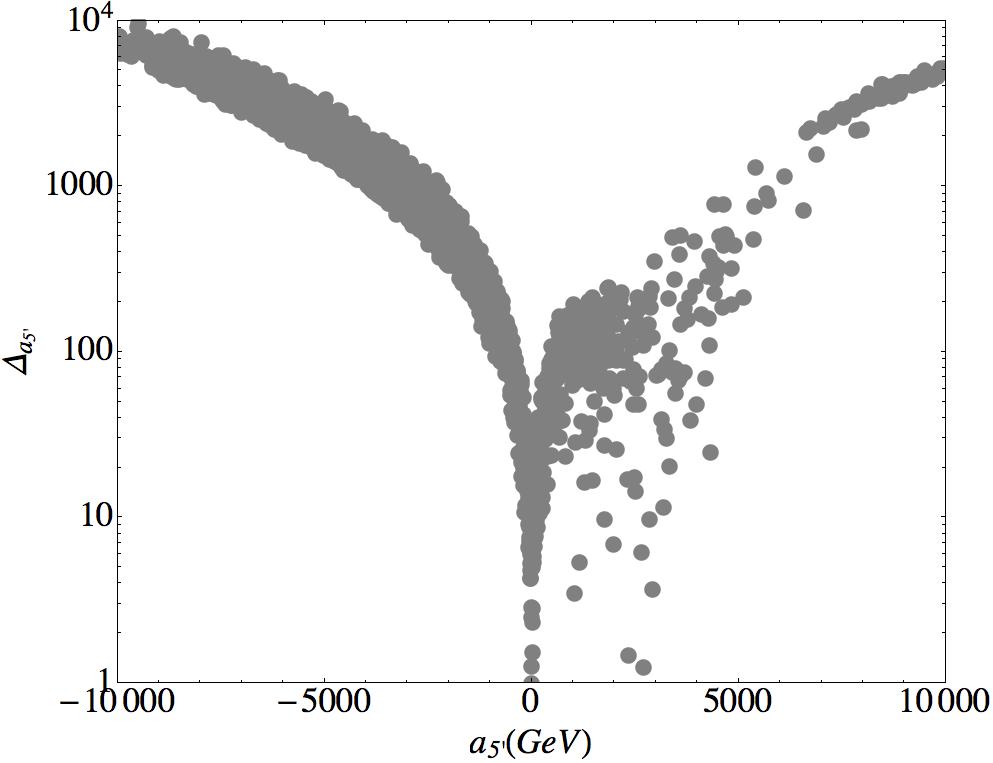}
\includegraphics[width=0.44\textwidth]{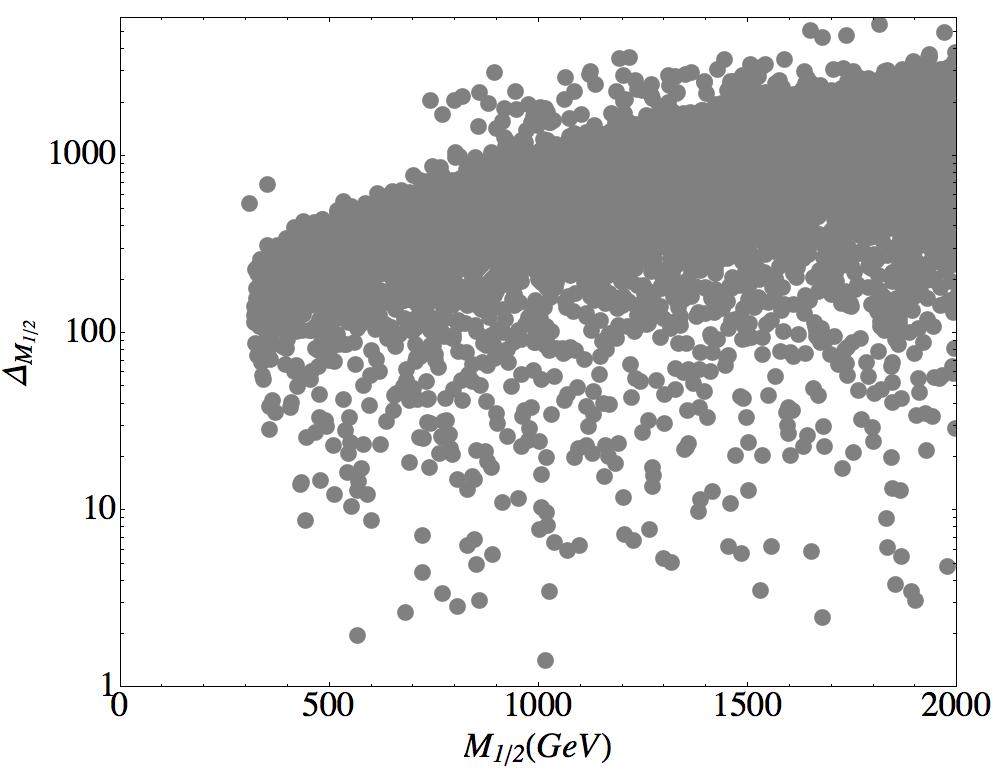}
\caption{\it Individual fine-tunings with respect to the input parameters $m_{10}$, $m_{5^{\prime}}$, $M_{1/2}$ and $a_{5^{\prime}}$ for non-universal gaugino mass scenarios.}
\label{fig:NUDeltas}
\end{figure}

This conjecture is supported by the individual fine tunings of $m_{10}$, $m_{5^{\prime}}$, $a_{5^{\prime}}$ and $M_{1/2}$ in Fig.~\ref{fig:NUDeltas}, where as before we see that in the limit of vanishing scalar masses the tuning tends to zero. This behaviour is in part due to the logarithmic form of the fine-tuning definition Eq.~(\ref{eq:FTdef}). The same behaviour is observed for the trilinear coupling but not for the gaugino mass $M_{1/2}$.  In contrast, for any value of the (GUT scale) gaugino mass, $M_{1/2}$, we find several points with no individual fine tuning of $M_{1/2}$. 

%%%%%%%%%%%%%%%%%
\subsection{An Enhanced Scan Over $M_{1/2}$, $\rho_1$ and $\rho_2$}

\begin{figure}[ht]
\centering
\includegraphics[width=0.48\textwidth]{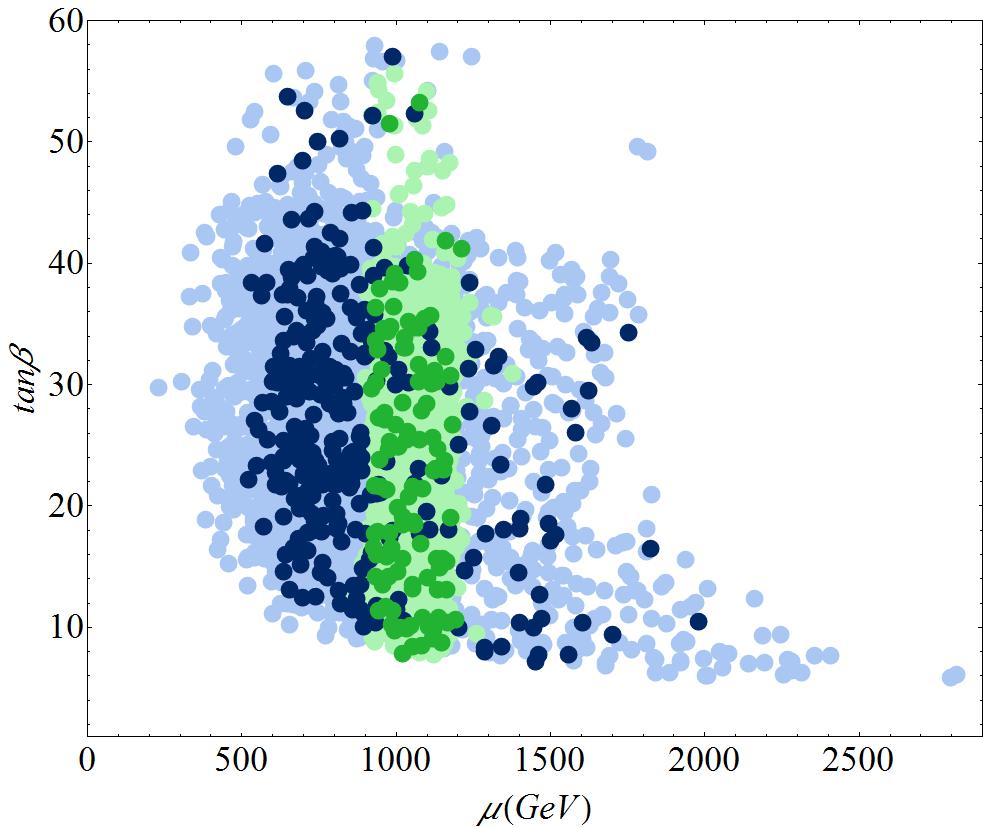}
\caption{\it Viable scenarios in the $\mu$-$\tan \beta$ plane for the enhanced scan with non-universal gaugino masses. Points with the preferred Dark Matter relic density are shown in green, while those with a relic density below the bounds are in blue. Darker and lighter shades denote the fine-tuning: darker shades have fine-tuning $\Delta<10$ while lighter shades have $10<\Delta<100$.}
\label{fig:NUmutanb_e}
\end{figure}

\begin{figure}[ht]
\centering
\includegraphics[width=0.48\textwidth]{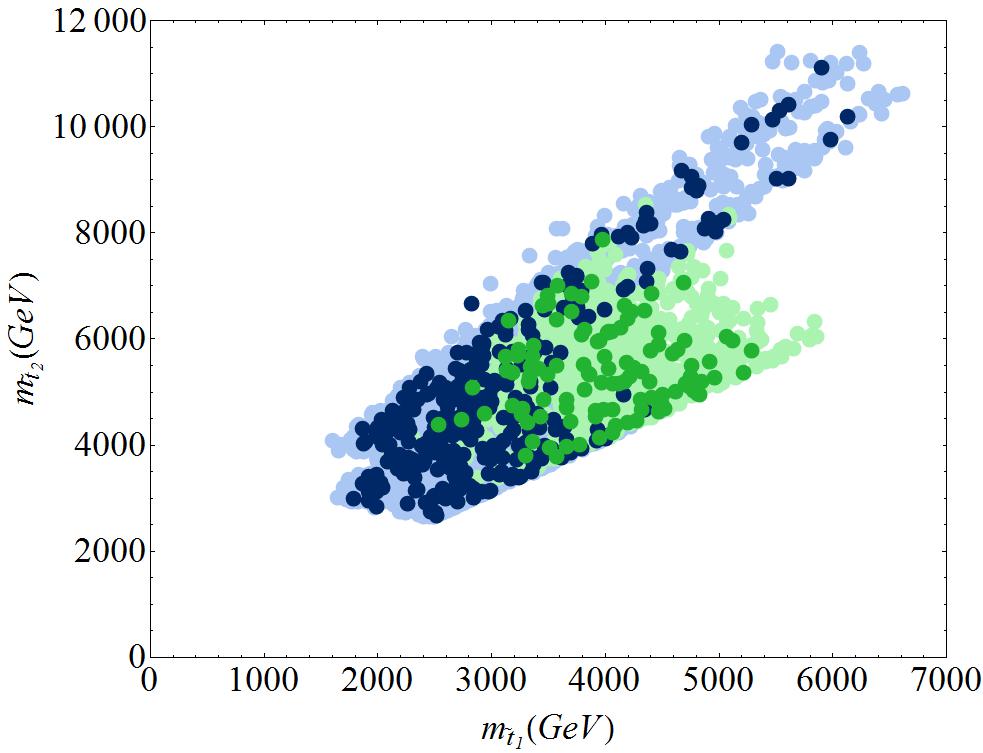}
\includegraphics[width=0.48\textwidth]{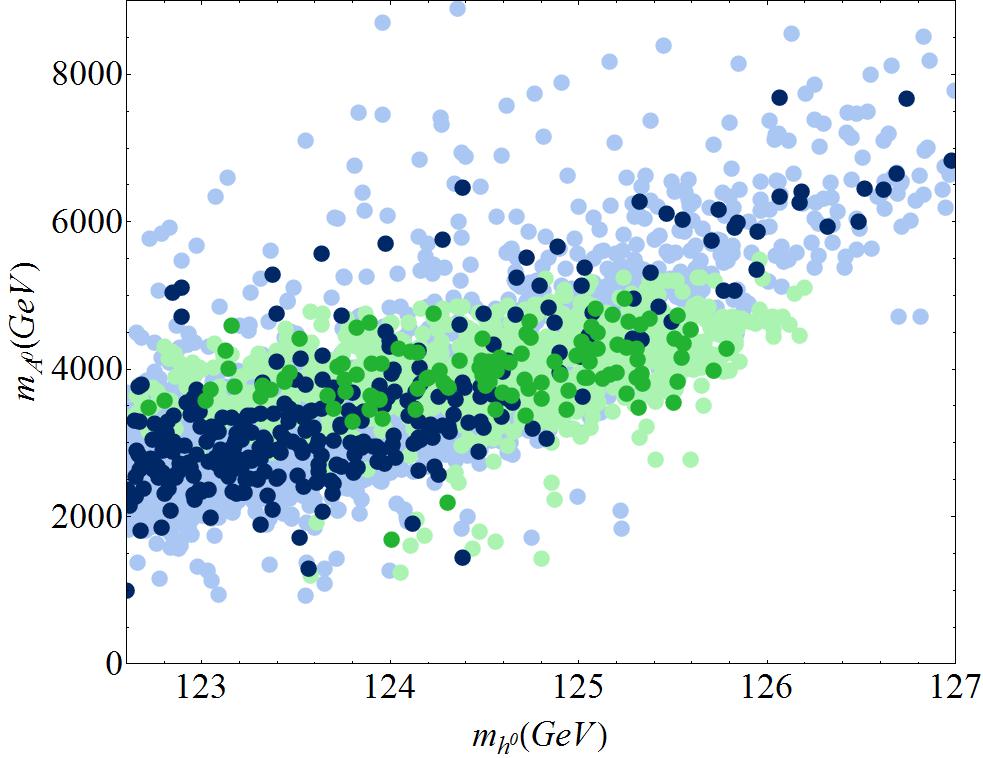}
\caption{\it Viable scenarios in the stop mass (left) and  lightest scalar - pseudoscalar mass (right) planes for the enhanced scan with non-universal gaugino masses, with colours as in Fig.~\ref{fig:NUmutanb_e}.}
\label{fig:NUstophiggs_e}
\end{figure}

To search for regions where the fine tuning of the soft parameters is small, we set the scalar masses and trilinear couplings to zero at the GUT scale\footnote{Setting these to be {\it exactly} zero is for computational simplicity only; any small value at the GUT scale should be overwhelmed by the large contribution from the gluino. In Secs.~\ref{sec:su5200} and \ref{sec:orbifolds} when we discuss explicity models we relax this and allow GUT scale scalar masses $<100\,{\rm GeV}$.}, but extend the range of the gaugino masses to \mbox{$0 < M_{1/2} < 5000\,{\rm GeV}$}. We allow $\rho_1$ and $\rho_2$ to vary over the interval $\left[-15,15\right]$, and only accept solutions where $\Delta < 100$ (again not including $\Delta_\mu$). Experimental and stability constraints are implemented as in the previous section. The surviving scenarios (3,832 out of approximately 130,000) are shown in the $\mu$-$\tan \beta$, stop mass and  Higgs mass planes in Fig.~\ref{fig:NUmutanb_e} and \ref{fig:NUstophiggs_e} and we now see not only points with fine-tuning less than 100 (lighter shades of green and blue) but also many with fine-tuning less than 10 (darker shades of green and blue). Furthermore, plenty points (1,028) provide a good description of the full Dark Matter relic density (green points) rather than describing only part of the relic density (blue points). 
We observe that insistence on the preferred dark matter abundance significantly restricts the allowed mass spectrum, and the preference for low fine-tuning narrows the allowed masses even further. In particular, for the optimal scenarios, we find $\mu$ restricted to be close to $1\,{\rm TeV}$, lightest top squarks confined to $2.5$-$5.5\,{\rm TeV}$ and the pseudoscalar Higgs boson mass around $4\,{\rm TeV}$. These ranges widen somewhat if we allow less dark matter or more fine-tuning. 

It is instructive at this point to discuss why some scenarios can provide such a low fine-tuning. Since we are neglecting fine-tuning from $\mu$, this is really a statement that $m_{H_u}$ is insensitive to fluctuation in the fundamental parameters. For the enhanced scan we have set the scalar masses and trilinears to zero, so the only dimensionful parameter that feeds the RGE's for $m_{H_u}$ is $M_{1/2}$ and at leading order one expects $m_{H_u}^2 = a M_{1/2}^2$ where $a$ is a dimensionless coefficient that depends only on the dimensionless parameters (such as the Yukawa couplings and $\rho_{1,2}$). Immediately this appears fine-tuned since a change in $M_{1/2}$ causes a proportionate change in $m_{H_u}$ . 

\begin{figure}[ht]
\centering
\includegraphics[width=0.48\textwidth]{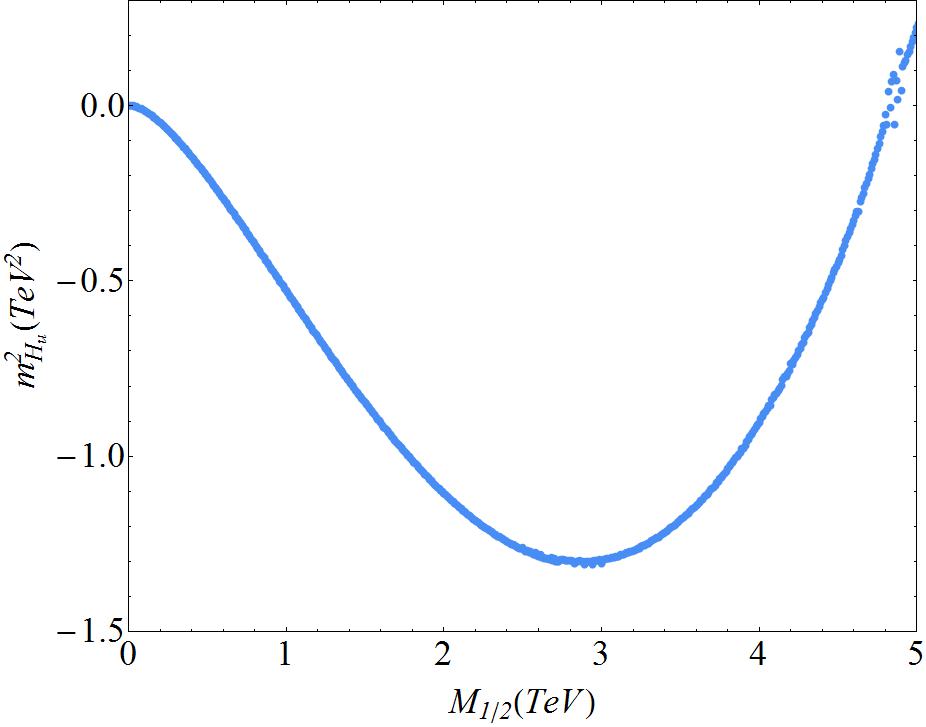}
\caption{\it The values of $m_{H_u}^2$ as $M_{1/2}$ is varied, for parameters as the BPO-I benchmark in Table~\ref{tab:Inputs} but with the scalar masses and trilinear couplings set to zero.}
\label{fig:mhu_m3}
\end{figure}

However, this expression is at leading order. One expects radiative corrections to electroweak symmetry breaking which 
are particularly important for the points on the ellipse, where $a$ is rather small. Taking these into account makes $a$ itself dependent on $M_{1/2}$ and a more complicated dependence results. This dependence on $M_{1/2}$ for typical parameters can be seen in Fig.~\ref{fig:mhu_m3}. In this particular case a choice of $M_{1/2} \approx 3\,$TeV sits close to a minimum, so $m_{H_u}^2$ is insensitive to fluctuations in $M_{1/2}$ while still having a large (absolute) value. 

\begin{figure}[ht!]
\centering
\includegraphics[width=0.48\textwidth]{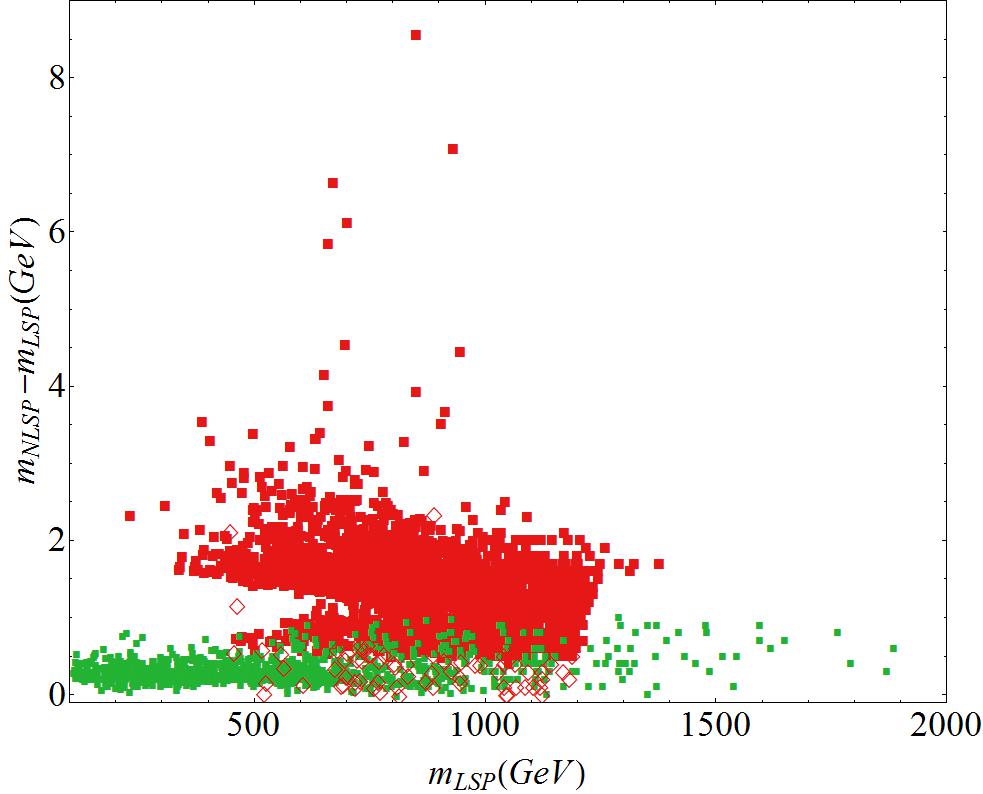}
\includegraphics[width=0.48\textwidth]{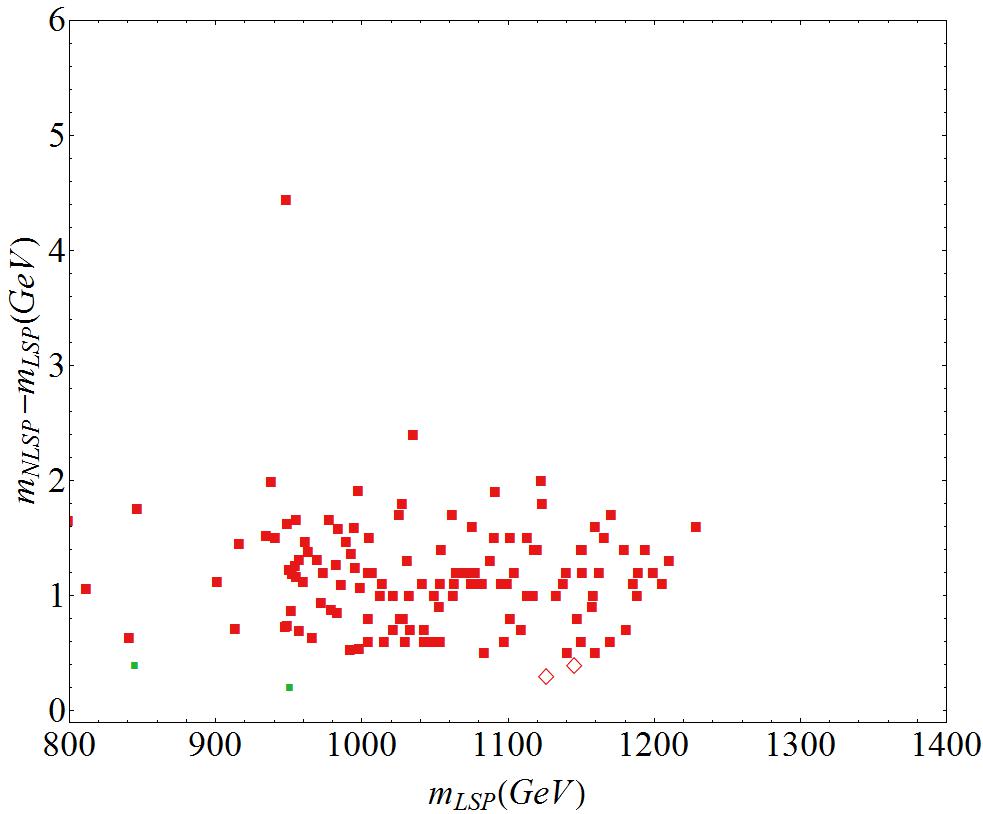}
\caption{\it Solutions in the plane of LSP mass vs.~the NLSP-LSP mass splitting for the enhanced scan over non-universal gaugino mass scenarios. The colour indicates the flavour of LSP, with red, blue and green denoting higgsino, bino and wino dominated Dark Matter respectively. The shape indicates the flavour of NLSP; filled squares and empty diamonds denote chargino and neutralino NLSP respectively. The left-hand plot shows all scenarios with fine-tuning $\Delta<100$ while the right-hand plot restricts to scenarios with $\Delta<10$ and the preferred Dark Matter relic abundance.}
\label{fig:NUDM_e}
\end{figure}

In Fig.~\ref{fig:NUDM_e}, we show the LSP and NLSP masses and nature. We see that these scenarios are as usual dominated by neutralino LSPs with chargino NSLPs but the relaxation of the gaugino universality now allows the LSP to be wino dominated. 

\begin{figure}[ht!]
\centering
\includegraphics[width=\textwidth]{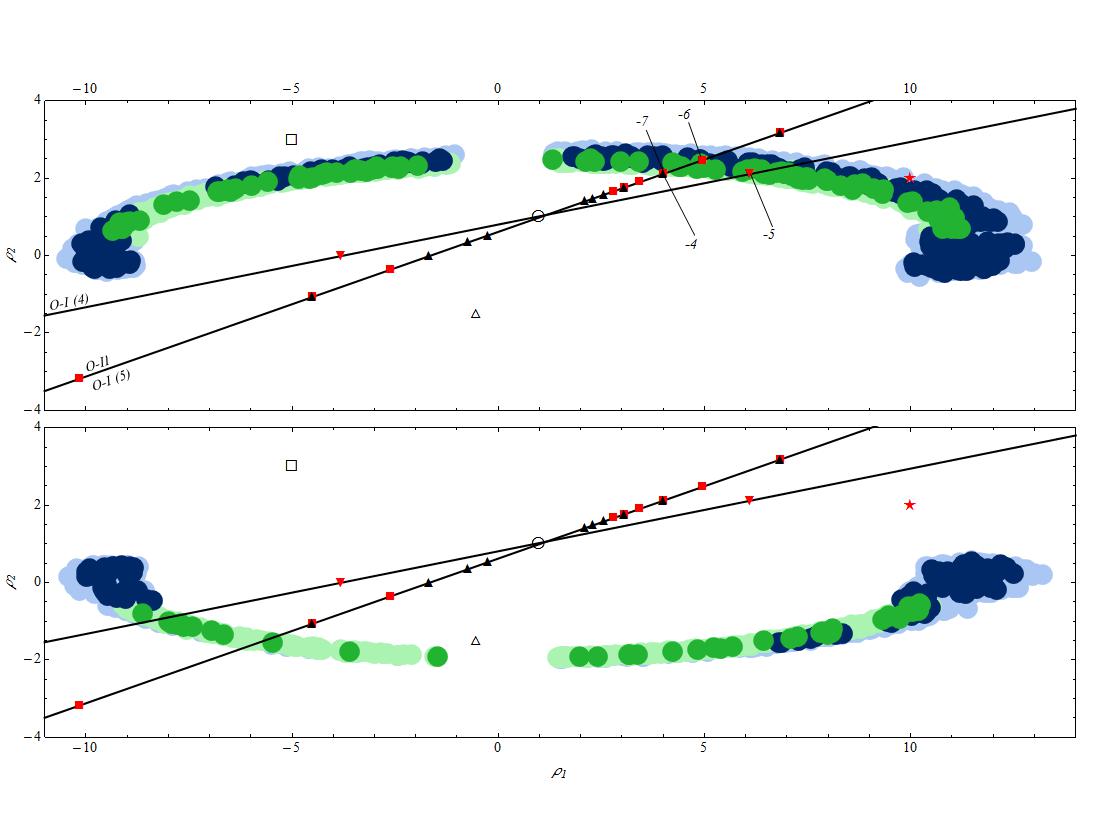}
\vspace*{-15mm} 
\caption{\it Viable scenarios in $\rho_1$-$\rho_2$ plane for the enhanced scan with non-universal gaugino masses. Points with the preferred Dark Matter relic density are shown in green, while those with a relic density below the bounds are in blue. Darker and lighter shades denote the fine-tuning: darker shades have fine-tuning $\Delta<10$ while lighter shades have $10<\Delta<100$. The upper pane is for scenarios with $\mu>0$ while the lower pane is for $\mu<0$. The additional symbols represent particular gaugino mass ratios as predicted by the mechanisms described in Sec.~\ref{sec:sfgmr}. Scenarios arising from embeddings in the $\mathbf{1}$, $\mathbf{24}$, $\mathbf{75}$, and $\mathbf{200}$ representations of SU(5) are shown by an empty circle, an empty triangle, an empty square and a red star respectively. The orbifold inspired scenarios lie along the straight lines: the O-I model with $n_{H} + n_{\overline{H}} = -4$ lies on the shallower gradient line while those for the  O-I model with $n_{H} + n_{\overline{H}} = -5$ share the steeper gradient line with the O-II orbifold. The numbers refer to $\delta_{GS}$ with those below the lines applicable to the O-I model and those above applicable to O-II.}
\label{fig:NUrho_e}
\end{figure}

Fig.~\ref{fig:NUrho_e} is divided into two panes, showing the $\rho_{1,2}$ values for positive and negative $\mu$ separately. We see that fine-tuning $<10$ favours positive values of $\mu$. It is interesting to note that all of these points fall on an ellipse. For $\mu>0$ ($\mu<0)$ the points on the bottom (top) half of the ellipse are excluded by our experimental constraint $P_{\rm tot}>10^{-3}$.  A similar analysis in Ref.~\cite{Antusch:2012gv} found a similar pattern. 

%%%%%%%%%%%%%%%%%%%%%%%%%%%%%%%%%%%%%%%%%%
\section{Scenarios with Fixed Gaugino Mass Ratios}
\label{sec:sfgmr}

In the above analysis we have implicitly assumed that the gaugino mass ratios are fixed by some GUT or string inspired mechanism. We here consider three classes of models as examples of how such mechanisms may be restricted by low energy constraints.  \\

\noindent {\bf 1.} The breaking of supersymmetry through a hidden sector field $\hat X$, with $f_{\alpha \beta}$ in a representation belonging to  the product $\left(\mathbf{24} \times \mathbf{24}\right)_{symm} = \mathbf{1} + \mathbf{24} + \mathbf{75} + \mathbf{200}$. The predicted gaugino mass ratios for embeddings in the $\mathbf{1}$, $\mathbf{24}$, $\mathbf{75}$, and $\mathbf{200}$ representations are shown in Fig.~\ref{fig:NUrho_e} by an empty circle, an empty triangle, an empty square and a red star respectively. \\

\noindent {\bf 2.} The Brignole, Ib\'a\~nez and Mu\~noz (BIM) O-I orbifold \cite{Brignole:1993dj} where the sum of the Higgs field modular weights\footnote{Here we use the notation adopted in \cite{Brignole:1993dj}.} are $n_{H} + n_{\overline{H}} = -5$ or $-4$. For simplicity we will consider here only moduli dominated scenarios\footnote{A dilaton dominated scenario would lie far from our ellipse.} with goldstino angles $ \theta =0$. Strictly speaking these models also restrict the scalar masses and force their mass-squared negative for $\sin^2 \theta \le 2/3$; here we disregard these scalar mass constraints and only use the orbifold to {\it inspire} values for the gaugino mass ratios. 
These ratios for the BIM O-I orbifold with $n_{H} + n_{\overline{H}} = -5$ are then given by\footnote{For the prefactors we use the values calculated in \cite{Brignole:1993dj}. This is a different approach from Refs.~\cite{Antusch:2012gv} and \cite{Horton:2009ed} where these coefficients are set to 1.}
\begin{eqnarray}
\rho_1 = 1.18 \,\frac{\delta_{GS} + 54/5}{\delta_{GS} + 6}, &\qquad&
\rho_2 = 1.06 \,\frac{\delta_{GS} + 8}{\delta_{GS} + 6}.
\label{eq:rhoO-I5}
\end{eqnarray}
$\delta_{GS}$ is a negative integer arising from the Green-Schwarz counterterm and required for anomaly cancellation. 
For $n_{H} + n_{\overline{H}} = -4$ they are
\begin{eqnarray}
\rho_1 = 1.18 \,\frac{\delta_{GS} + 51/5}{\delta_{GS} + 6}, &\qquad&
\rho_2 = 1.06 \,\frac{\delta_{GS} + 7}{\delta_{GS} + 6}.
\label{eq:rhoO-I4}
\end{eqnarray}
These scenarios are represented in Fig.~\ref{fig:NUrho_e} by filled black triangles triangles and inverted red triangles respectively. Note that each of these orbifold models provide scenarios that lie along a line in the $\rho_1$-$\rho_2$ plane (also drawn in Fig.~\ref{fig:NUrho_e}).\\

\noindent {\bf 3.} The BIM O-II orbifold for which 
\begin{eqnarray}
\rho_1 = 1.18 \,\frac{b_1 - \delta_{GS}}{b_3 - \delta_{GS}}, &\qquad&
\rho_2 = 1.06 \,\frac{b_2 - \delta_{GS}}{b_3 - \delta_{GS}}.
\label{eq:rhoO-II}
\end{eqnarray}
$b_{1,2,3}= \left(33/5, 1, -3 \right)$ are the usual MSSM one-loop beta function coefficients. Again we are assuming moduli domination and neglecting the scalar mass predictions. These models share the line of the O-I models with $n_{H} + n_{\overline{H}} = -5$ in the $\rho_1$-$\rho_2$ plane and are identified in Fig.~\ref{fig:NUrho_e} by filled red squares.  \\

We observe that only five models provide mass ratios that lie close to our ellipse: hidden sector breaking with a $\mathbf{200}$; the BIM O-I orbifold with $n_{H} + n_{\overline{H}} = -4$ and $\delta_{GS} = -5$; the BIM O-I orbifold with $n_{H} + n_{\overline{H}} = -5$ and $\delta_{GS} = -4$ which coincides with the BIM O-II orbifold with $\delta_{GS} = -7$; and the BIM O-II orbifold with $\delta_{GS} = -6$. All of these models coincide with the upper half of the ellipse, so require ${\rm sign}(\mu) = +$.  We will now study these cases individually.

%%%%%%%%%%%%%%%%%%%%%%%
\subsection{$SU(5)_{\mathbf{200}}$ Model}
\label{sec:su5200}

We first consider the model with a gauge-kinetic function embedded in a $\mathbf{200}$ of $SU(5)$, generating the GUT scale gaugino mass ratios $\rho_1 = 10$ and $\rho_2 = 2$. We note in advance that this model lies towards the edge of the ellipse in Fig.~\ref{fig:NUrho_e}, in a light blue region, indicating that it may be difficult to generate points with small fine-tuning.  When we perform a detailed scan we find that this is indeed the case; all viable scenarios have $\Delta \gtrsim 75$. 
However, despite its unattractive fine-tuning, this model also provides some predictions. 

\begin{figure}[h!]
\centering
\includegraphics[width=0.48\textwidth]{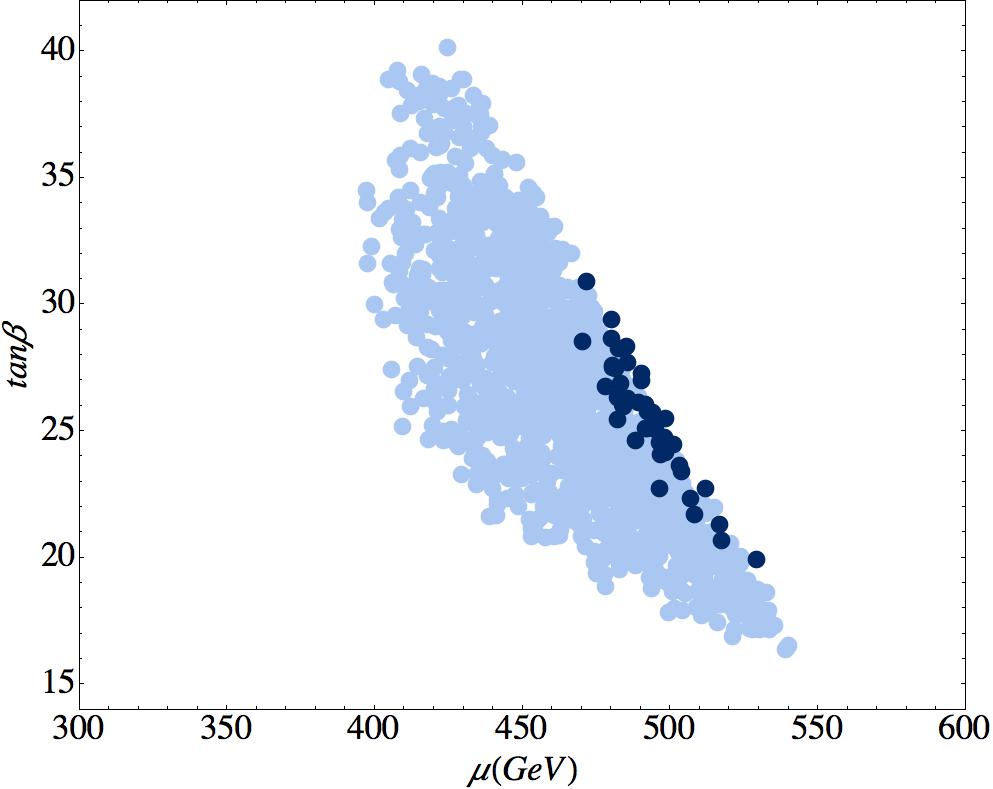} 
\caption{\it Viable scenarios in the $\mu$-$\tan \beta$ plane for the $SU(5)_\mathbf{200}$ model. Darker and lighter shades denote the fine-tuning: darker shades have fine-tuning $\Delta<80$ while lighter shades have \mbox{$80<\Delta<100$}. All these scenarios have a Dark Matter relic density below the preferred range.}
\label{fig:200mutanb}
\end{figure}

\begin{figure}[h!]
\centering
\includegraphics[width=0.48\textwidth]{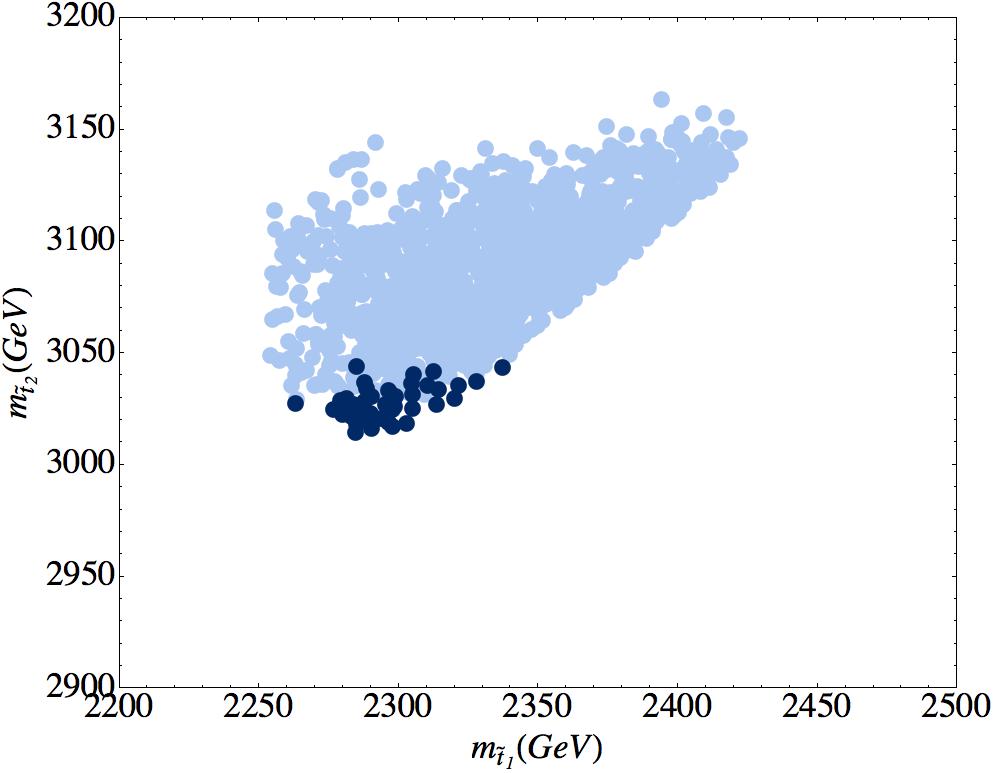}
\includegraphics[width=0.48\textwidth]{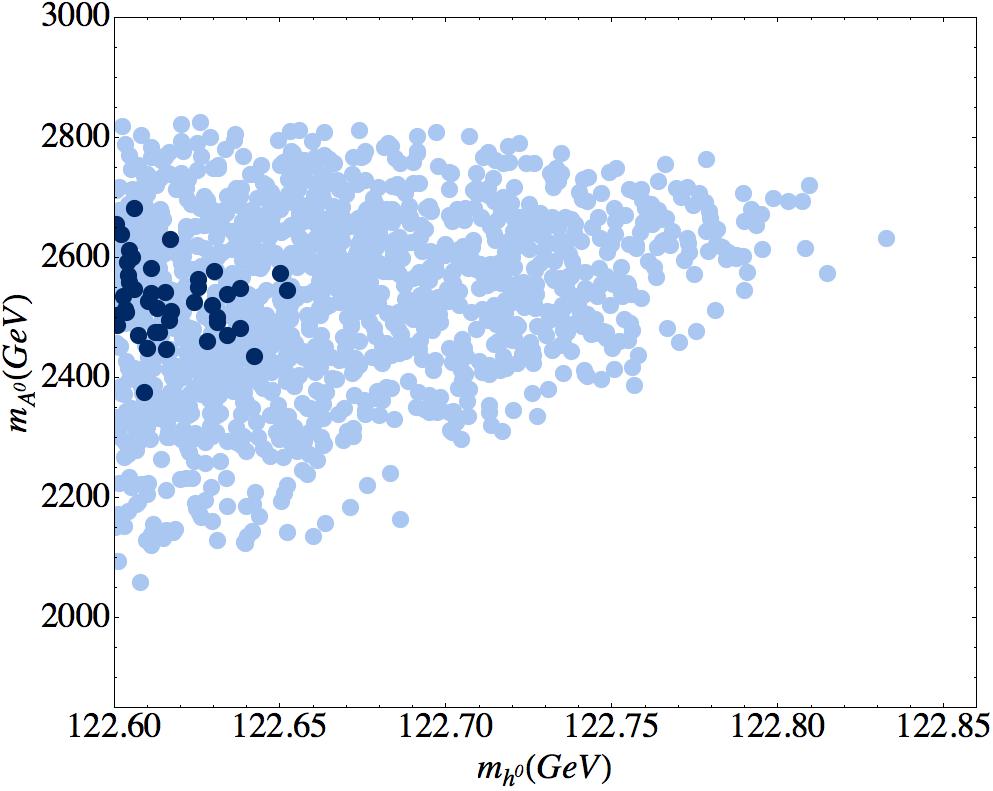}
\caption{\it Viable scenarios in the stop mass (left) and  lightest scalar - pseudoscalar mass (right) planes for the $SU(5)_\mathbf{200}$ model, with colours as in Fig.~\ref{fig:200mutanb}.}
\label{fig:200stophiggs}
\end{figure}

Firstly, the value of $\tan \beta$ is quite large, see Fig.~\ref{fig:200mutanb}, in the range $16-41$, and this becomes more resticted\footnote{Note that the definition of dark and light shades in Figs.~\ref{fig:200mutanb} and 
\ref{fig:200stophiggs} differ from those of Figs.~\ref{fig:Umutanb}, \ref{fig:Ustophiggs},
\ref{fig:NUmutanb} and \ref{fig:NUstophiggs} since we have no scenarios with $\Delta < 10$.}, $20 - 32$, if we insist that $\Delta < 80$. This favours $\mu \sim 500\,{\rm GeV}$ with a corresponding higgsino-dominated neutralino as Dark Matter. Unfortunately, this contributes only $\sim 30\%$ of the preferred relic density, but unlike the sfermions, it should be within reach of the $14\,{\rm TeV}$ LHC. 

The allowed stop, sbottom and stau masses as well as the Higgs mass are also restricted to rather small regions of parameter space for viable scenarios. Scenarios in the stop and Higgs mass planes are shown in Fig.~\ref{fig:200stophiggs}. The lightest stop has a mass of around $2.25$ - $2.35\, {\rm TeV}$ for $\Delta<80$. We see similar restrictions for the sbottom and stau, but do not reproduce the plots here (see Tab.~\ref{tab:1st3rd} for two typical scenarios). These are probably outside the reach of the $14\,{\rm TeV}$ LHC. It is also rather difficult to keep the Higgs mass heavy  with $m_{h^0} \sim 122.6$ GeV for all solutions with $\Delta < 80$, though this is still compatible with the current combined experimental and theoretical uncertainties.

The LSP in this scenario is  exclusively a higgsino dominated neutralino with mass that closely follows the value of $\mu$. The NLSP is similarly a higgsino dominated chargino, always between $1$ - $2\,{\rm GeV}$ heavier. Note that for these scenarios $\Delta_{\mu} \sim 120$ so comparable with the other fine-tunings. 

%%%%%%%%%%%%%%%%%%%%%
\subsection{BIM Orbifold Models}
\label{sec:orbifolds}

The BIM O-I orbifold with $n_{H} + n_{\overline{H}} = -5$ and $\delta_{GS} = -7$ (and the coincident BIM O-II orbifold with $\delta_{GS} = -7$) also lies towards the edge of the ellipse with $\rho_1 = 4.01$ and $\rho_2 = 2.12$. However, in this case a dedicated scan finds no viable scenarios since the Dark Matter relic density is always too large (by about a factor of seven). Therefore we conclude that this model with $\Delta<100$ is already ruled out. 

The BIM O-I orbifold with $n_{H} + n_{\overline{H}} = -4$ and $\delta_{GS} = -5$ predicts $\rho_1 = 6.14$ and $\rho_2 = 2.12$.  This lies very near the ellipse of Fig.~\ref{fig:NUrho_e} and when we perform a dedicated scan over its parameter space, we do indeed find plenty of solutions with low fine tuning. Rather intriguingly the majority of our points have a Dark Matter relic density in the preferred range. It is quite remarkable that this model agrees so well with all low energy data while still allowing (non-$\mu$) fine-tuning to be very small. 

\begin{figure}[h!]
\centering
\includegraphics[width=0.48\textwidth]{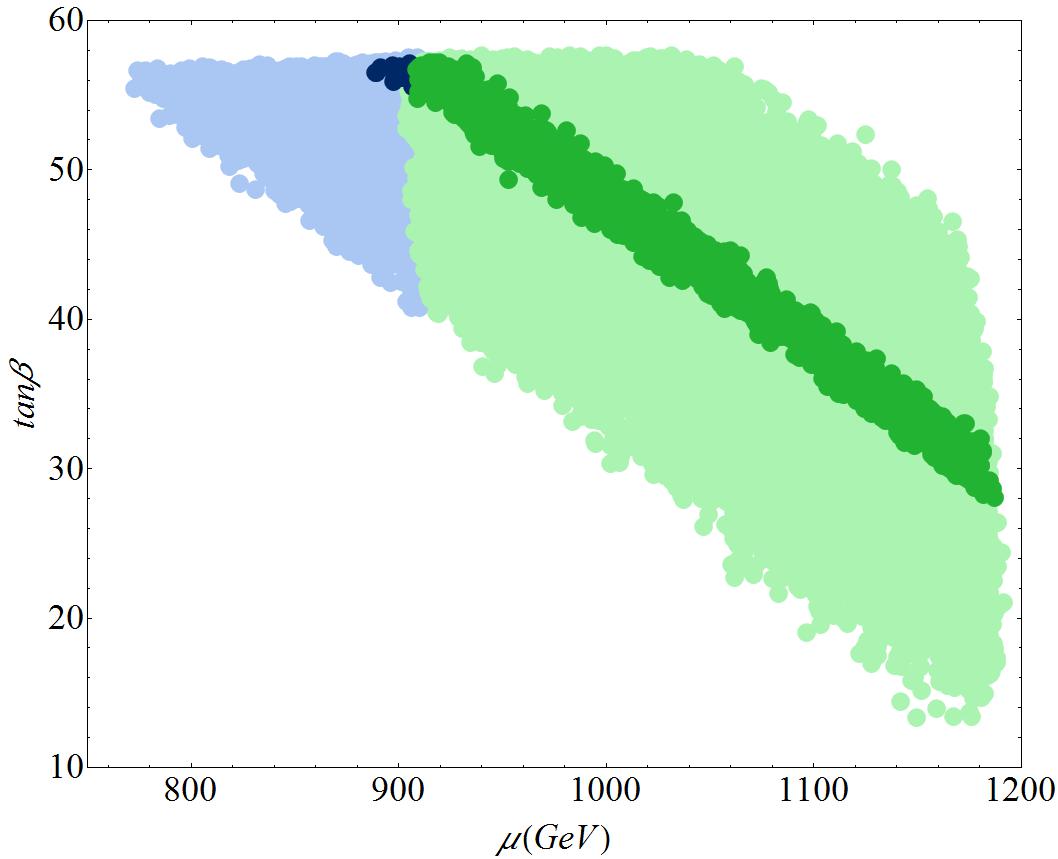} 
\caption{\it Viable scenarios in the $\mu$-$\tan \beta$ plane for the O-I orbifold model with $\delta_{GS} = -5$. All points have the preferred Dark Matter relic density. Darker and lighter shades denote the fine-tuning: darker shades have fine-tuning $\Delta<10$ while lighter shades have $10<\Delta<100$.}
\label{fig:O-I5mutanb}
\end{figure}

In Fig.~\ref{fig:O-I5mutanb} we show the values of $\mu$ and $\tan \beta$ for the viable scenarios, indicating a preference for moderate to large values of $\tan \beta$, between 28 and 58 for fine-tuning $\Delta < 10$.  $\mu$ is now necessarily quite large, around $0.9-1.2\,{\rm TeV}$ for the least fine-tuned scenarios; lower values of $\mu$ produce an insufficient Dark Matter relic density. The LSP (neutralino) and NLSP (chargino) are both higgsino dominated and lie within roughly $1\,{\rm GeV}$ of each other. The distinct upper bound on $\tan \beta$ is due to our requirement for vacuum stability, while the distinct upper bound on $\mu$ is due to the upper bound on the Dark Matter relic density. The diagonal boundaries are caused by our fine-tuning constraint.

\begin{figure}[h!]
\centering
\includegraphics[width=0.48\textwidth]{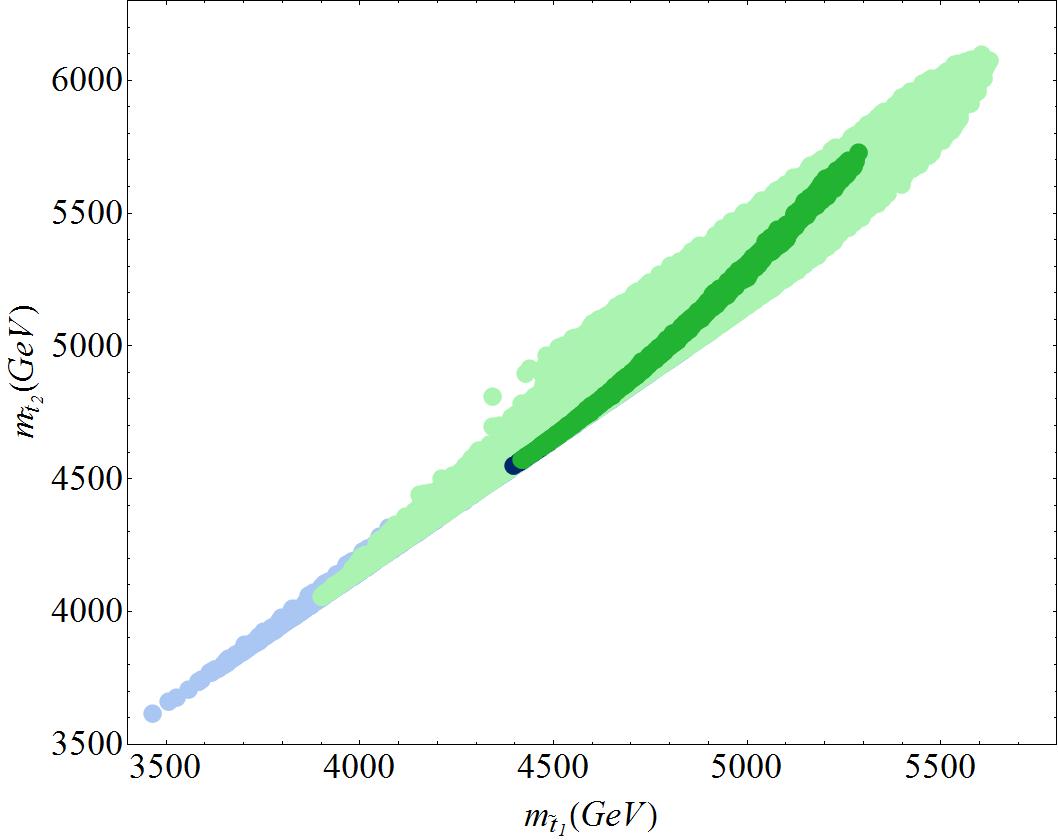}
\includegraphics[width=0.48\textwidth]{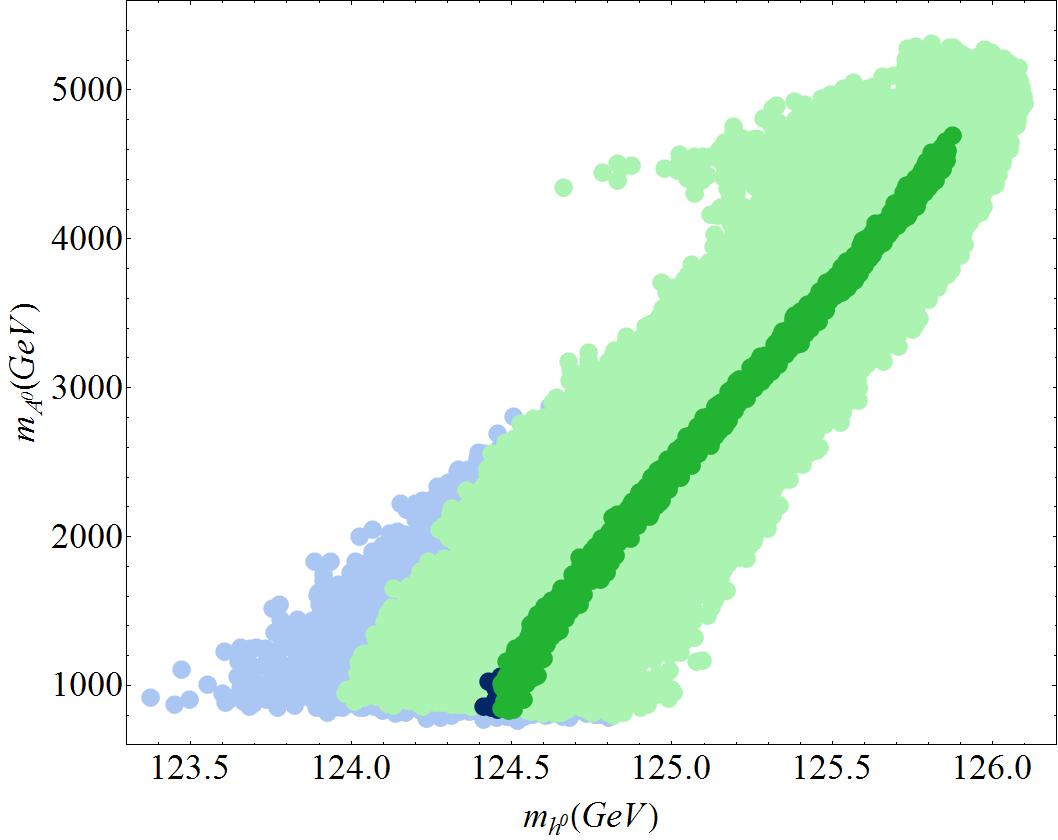}
\caption{\it Viable scenarios in the stop mass (left) and  lightest scalar - pseudoscalar mass (right) planes for the O-I orbifold model with $\delta_{GS} = -5$, with colours as in Fig.~\ref{fig:O-I5mutanb}.}
\label{fig:O-I5stophiggs}
\end{figure}

The stop masses and Higgs masses are shown in Fig.~\ref{fig:O-I5stophiggs}. 
Now we have really very heavy stops, which in turn contribute to the Higgs mass radiative corrections, making it much easier to obtain the correct Higgs mass. Indeed once the other constraints are applied these models seem to prefer a lightest scalar Higgs between $124.5$ and $126\,{\rm GeV}$. Since the pseudoscalar mass is now very heavy, this lightest scalar would look exactly like the SM Higgs boson, in accordance with the most recent findings. \\

\begin{figure}[h!]
\centering
\includegraphics[width=0.48\textwidth]{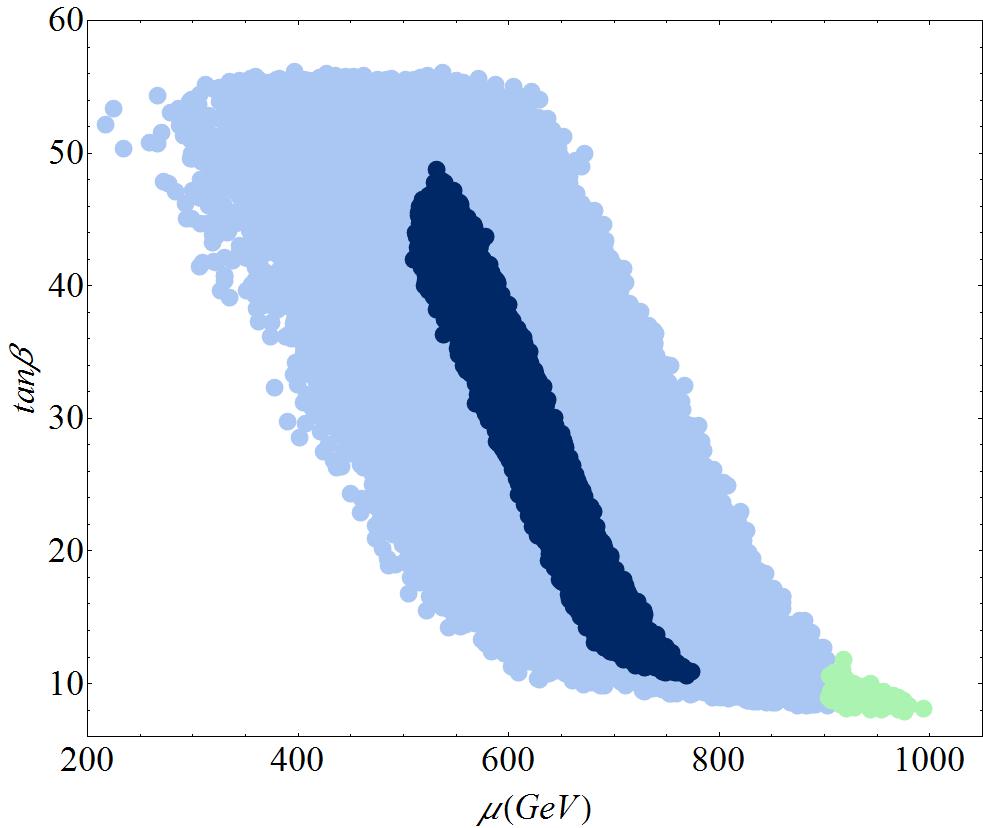}
\caption{\it Viable scenarios in the $\mu$-$\tan \beta$ plane for the O-II orbifold model with $\delta_{GS} = -6$. Points with the preferred Dark Matter relic density are shown in green, while those with a relic density below the bounds are in blue. Darker and lighter shades denote the fine-tuning: darker shades have fine-tuning $\Delta<10$ while lighter shades have $10<\Delta<100$.}
\label{fig:O-IImutanb}
\end{figure}

The only other BIM orbifold that lies on the ellipse of Fig.~\ref{fig:NUrho_e} is the O-II orbifold with $\delta_{GS} = -6$. This predicts $\rho_1 = 4.96$ and $\rho_2 = 2.47$. The viable scenarios in the $\mu-\tan\beta$ plane are shown in Fig.~\ref{fig:O-IImutanb}. Now we see that most points have a Dark Matter relic density that lies below the preferred range, though we now have more moderate values of $\mu$ allowed, as low as about $200\,{\rm GeV}$. These low $\mu$ points still have fine-tuning of order $\Delta \sim 100$ from the other parameters, so having $\mu$ small gains us nothing in this regard. To keep $\Delta \le 10$ requires $\mu$ larger than about $500\,{\rm GeV}$. The scenarios with the preferred relic density all fall in the tail of the distribution, with quite low values of $\tan \beta$ and have fine-tuning $\Delta \sim 100$.

\begin{figure}[h!]
\centering
\includegraphics[width=0.48\textwidth]{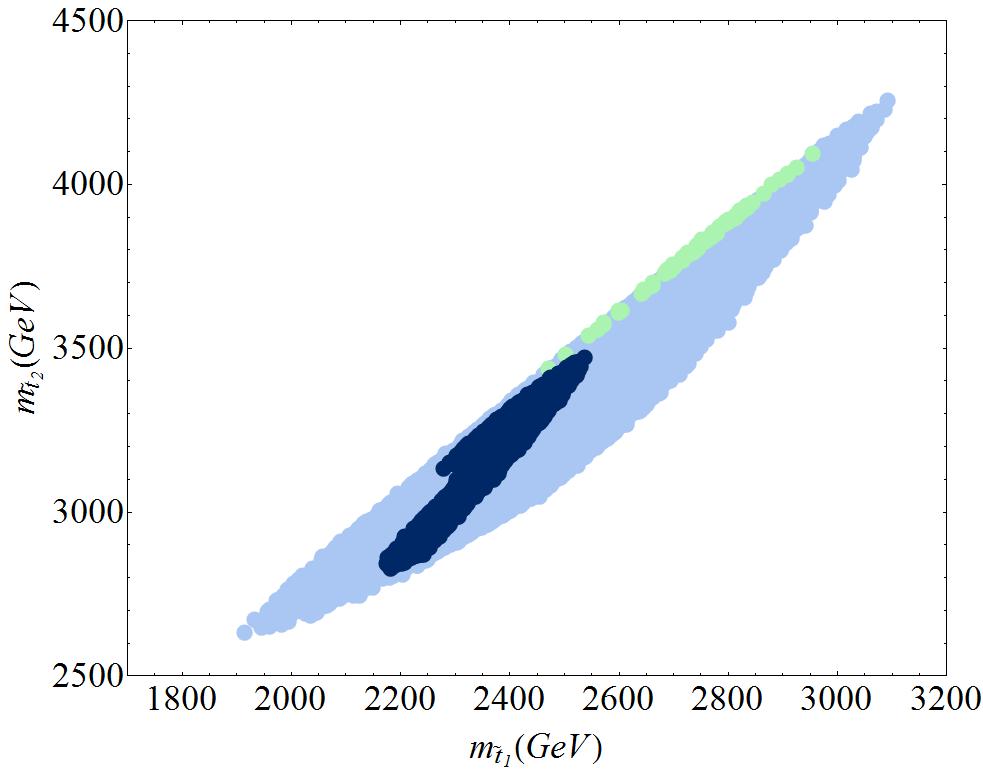}
\includegraphics[width=0.48\textwidth]{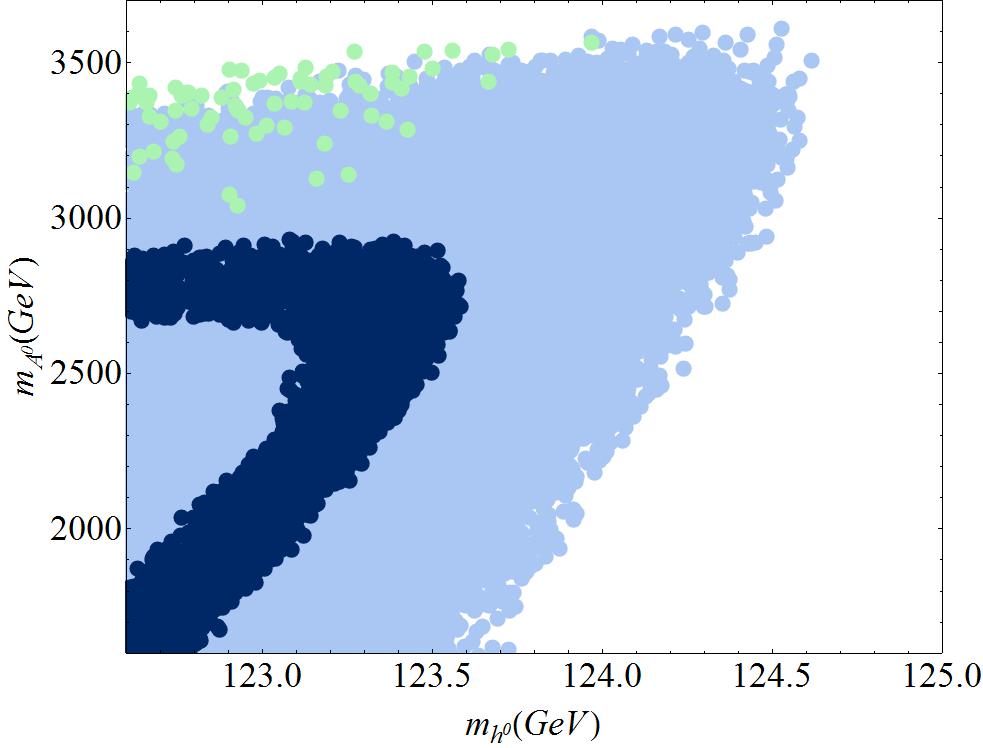}
\caption{\it Viable scenarios in the stop mass (left) and  lightest scalar - pseudoscalar mass (right) planes for the O-II orbifold model with $\delta_{GS} = -6$, with colours as in Fig.(\ref{fig:O-IImutanb}).}
\label{fig:O-IIstophiggs}
\end{figure}

In Fig.~\ref{fig:O-IIstophiggs} we show the results obtained in the stop mass and Higgs mass planes. The stops are considerably lighter than in the previous O-I example, even for scenarios with the preferred relic density, making them more attractive for LHC searches (though still very challenging). The corollary of lighter stops is that we also have a lighter Higgs boson, though as for the $SU(5)_{\mathbf{200}}$ this does not exclude the scenarios. 
Once again, both the LSP and NLSP are higgsino dominated (neutralino and chargino respectively) and separated by about a {\rm GeV}. 

%%%%%%%%%%%%%%%%%%%%%%%%%%%%%%%%%
\subsection{First and Second Generation Squarks and Gluinos}

\begin{figure}[h!]
\centering
\includegraphics[width=0.48\textwidth]{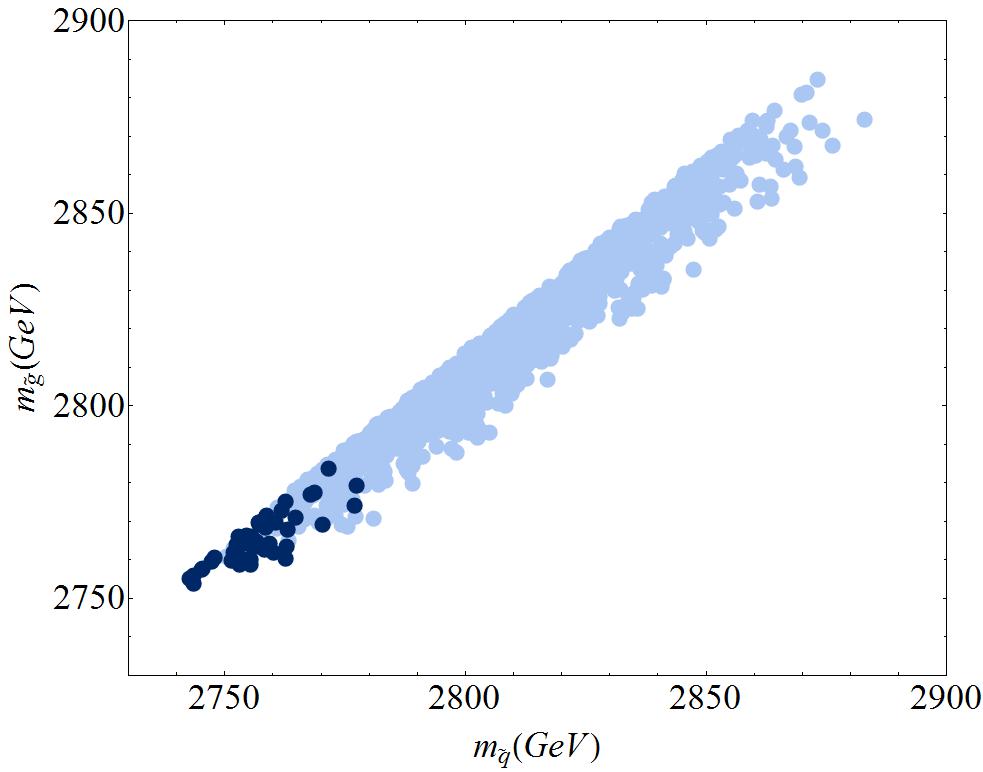}\\
\includegraphics[width=0.48\textwidth]{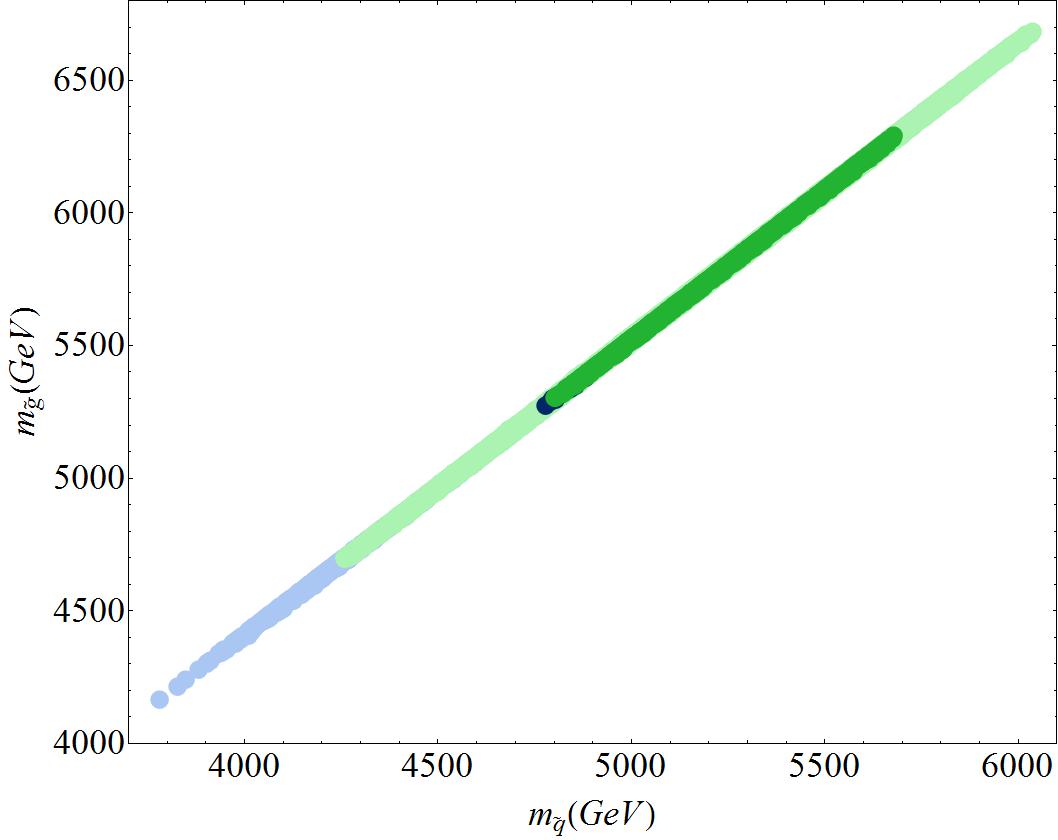}
\includegraphics[width=0.48\textwidth]{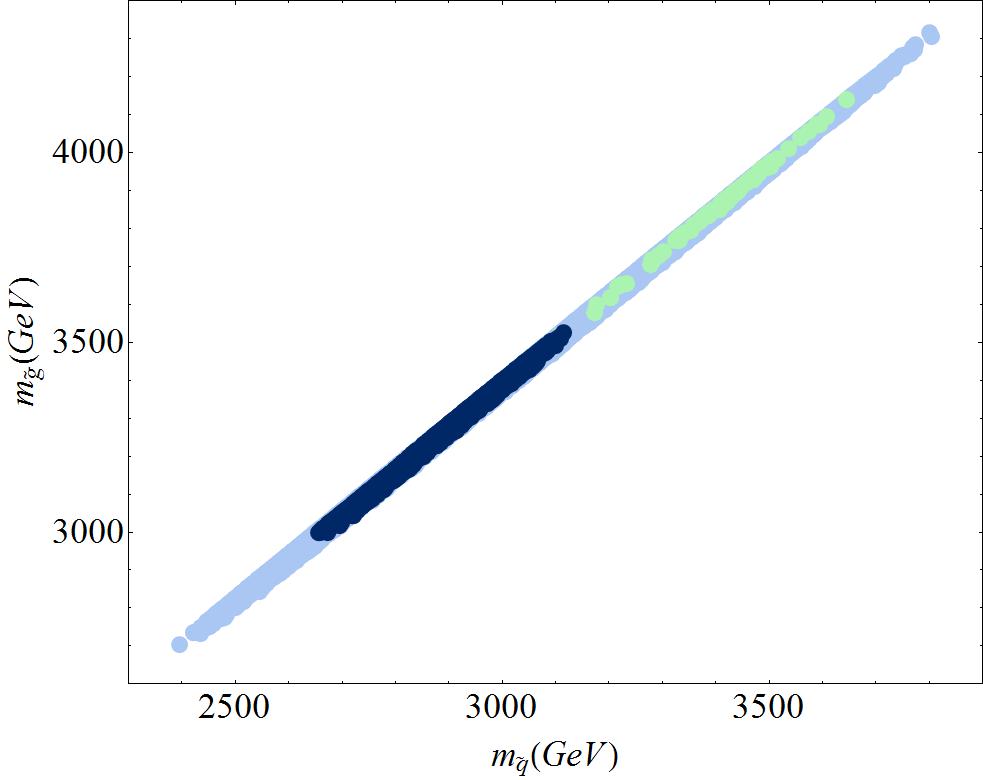}
\caption{\it The lightest squark mass and the gluino mass for the $SU(5)_\mathbf{200}$ model (top), the O-I orbifold model with $\delta_{GS} = -5$ (bottom-left) and O-II orbifold model with $\delta_{GS} = -6$ (bottom-right). Points with the preferred Dark Matter relic density are shown in green, while those with a relic density below the bounds are in blue. Darker and lighter shades denote the fine-tuning: in the upper plot ($SU(5)_\mathbf{200}$ scenarios), darker shades have fine-tuning $\Delta<80$ while lighter shades have $80<\Delta<100$; in the two lower plots (orbifold scenarios) darker shades have fine-tuning $\Delta<10$ while lighter shades have $10<\Delta<100$}
\label{fig:gluino}
\end{figure}

The masses of gluinos and first and second generation squarks are important for the potential discovery of supersymmetry  \cite{BrennerMariotto:2008gd}. In Fig.~\ref{fig:gluino}, we show the gluino mass $m_{\tilde{g}}$ in comparison to the lightest squark mass for the three models with viable scenarios discussed in Secs.~\ref{sec:su5200} and \ref{sec:orbifolds}. In all three cases we see a striking correlation between the gluino mass and the lighest squark mass. This can be easily understood analytically by making some simplifying approximations. 

It is well known that the one-loop RGEs for the soft gaugino masses $M_i$ are analytically solvable, giving
\begin{equation}
M_i(t) = M_i(0) \frac{\alpha_i(t)}{\alpha_i(0)}.
\label{eq:gaugino}
\end{equation}
Similarly, when one neglects the small Yukawa couplings, the one-loop RGEs for the first and second generation squarks are also analytically solvable (see Ref.~\cite{Miller:2012vn} for a discussion), giving
\begin{equation}
m^2_{\tilde d_R}(t) = m^2_{\tilde d_R}(0) 
-  \frac{8}{9} M_3^2 (0) \left[ \frac{ \alpha_3^2(0)-\alpha_3^2(t)}{\alpha_3^2(0)} \right]
+ \frac{2}{99} M_1^2 (0) \left[ \frac{ \alpha_1^2(0)-\alpha_1^2(t)}{\alpha_1^2(0)} \right],
\end{equation}
where we use the $\tilde d_R$ squark mass as an example, and ignore the contribution from the Higgs soft scalar masses which is always small for these scenarios. Using $M_1(0) = M_3(0) \rho_1$, applying Eq.~(\ref{eq:gaugino}), using the boundary condition
$m^2_{\tilde d_R}(0) = K_{\mathbf{\overline{5}}}  m^2_{\mathbf{\overline{5}}}$ and putting in numbers for the couplings, this gives approximately
\begin{equation}
m^2_{\tilde d_R}(t) = K_{\mathbf{\overline{5}}}  m^2_{\mathbf{\overline{5}}}
+ M_3^2 (t) \left[ 0.78 + 0.002 \,\rho_1^2 \right].
\label{eq:approxdr}
\end{equation}
When $m_{\tilde d_R}(0) $ is kept small, the dominant contribution arises from the gluino mass. For the two orbifold models, $\rho_1$ is also rather small so one has $m_{\tilde d_R} \approx 0.9 \,m_{\tilde g}$. For the $SU(5)_\mathbf{200}$ scenarios, the larger $U(1)$ gaugino mass ($\rho_1=10$) pushes this up a little to give $m_{\tilde d_R} \approx \,m_{\tilde g}$. The small spread in squark masses for a particular gluon mass is mainly caused by variations in $K_{\mathbf{\overline{5}}}$. Note that the apparent greater spread in masses for the $SU(5)_\mathbf{200}$ scenarios in Fig.~\ref{fig:gluino} is only due to the different plot scales.
Eq.~(\ref{eq:approxdr}) actually also works for the first two scenarios in Tab.~\ref{tab:Inputs} because coincidentally these scenarios have $m^2_{\mathbf{5}^\prime} \approx m^2_{\mathbf{\overline{5}}^\prime}$ so that their contributions cancel, but does not work in general. Of course this argument is very approximate and ignores all the extra contributions that are included in the full two-loop SOFTSUSY analysis but nevertheless gives good qualitative agreement.

It is interesting that the  $SU(5)_\mathbf{200}$ scenarios all require gluino and lightest squark masses in a rather restricted window, ranging from about $2740\, {\rm GeV}$ to about $2890\,{\rm GeV}$, so well beyond current LHC limits. Requiring $\Delta <80$ restricts them further to the very start of this already narrow mass window. If this model is a true reflection of reality, it is not surprising that the LHC has not yet seen supersymmetry. However, such gluino masses should be observable at the $14\,{\rm TeV}$ LHC. 

The orbifold models also restrict the gluino and lightest squark masses but the window is much larger.  For the O-I model we find viable scenarios only with the lightest squarks heavier than about $3.7\,{\rm TeV}$ and the gluinos about 10\% heavier. Requiring $\Delta <10$ results in the lightest squark being heavier than about $4.8\,{\rm TeV}$. Unfortunately these scenarios are considerably beyond the expected reach of  the $14\,{\rm TeV}$ LHC~\cite{Baer:2009dn}, which is unfortunate since this is our most attractive possibility, able to explain the entirety of Dark Matter while simultaneously keep the fine-tuning in the soft mass parameters small. Nevertheless, an energy-upgraded Super-LHC with $\sqrt{s} = 28\,{\rm TeV}$ would enhance production rates of such squarks and gluinos by a factor of ten~\cite{Bruning:2002yh}, allowing these scenarios to become accessible. 

The O-II model is also restrictive, but like the $SU(5)_\mathbf{200}$ scenarios allows squarks and gluinos within reach of the $14\,{\rm TeV}$ LHC. If fine-tuning is our priority then we may achieve $\Delta <10$ with lightest squark masses between about $2.6\,{\rm TeV}$ and $3.1\, {\rm TeV}$, but if the preferred Dark Matter relic density is desired one requires a slightly heavier lightest squark between about $3.2\,{\rm TeV}$ and $3.7\, {\rm TeV}$. Unfortunately this models does not allow low fine-tuning and the preferred relic density simultaneously. 

%%%%%%%%%%%%%%%%%%%%%%%%%
\section{Benchmark Points}\label{sec:BP}

In this section we present six benchmarks for viable $SU(5)$ GUT scenarios with non-universal masses that may be interesting to consider at either the $14\,{\rm TeV}$ LHC or the energy-upgraded Super-LHC with \mbox{$\sqrt{s} = 28\,{\rm TeV}$}.  The GUT scale parameters for these scenarios can be found in Tab.~\ref{tab:Inputs}. In Tab.~\ref{tab:Higgs} we show the masses of the five Higgs bosons. The masses of the first and third generation sfermions are shown in Tab.~\ref{tab:1st3rd}. The second generation sfermions are assumed degenerate with the first. In Tab.~\ref{tab:gaugino} we show the gaugino masses. Finally in Tab.~\ref{tab:pars} we present $\mu$, $\tan \beta$, the fine-tuning $\Delta$, the fine-tuning from $\mu$ alone, the predicted relic density of Dark Matter, and the predominant component of the LSP.

\begin{table}[h!]
\centering
\begin{tabular}{>{\centering}m{1.cm}>{\centering}m{2.cm}>{\centering}m{2.cm}>{\centering}m{2.25cm}>{\centering}m{2.25cm}
>{\centering}m{1.8cm}>{\centering}m{1.8cm}}
 & $\rm{BP1}SU(5)_{\mathbf{1}}$ & $\rm{BP2}SU(5)_{\mathbf{1}}$ &  $\rm{BP1}SU(5)_{\mathbf{200}}$ &  $\rm{BP2}SU(5)_{\mathbf{200}}$ & BPO-I & BPO-II \tabularnewline
\hline \\[-4mm]
%\hline \hline
$m_{\mathbf{10}}$  & 3305 & 2632 & 78.86 & 70.97 & 9.33 & 24.75 \tabularnewline
%\hline
$m_{\mathbf{\overline{5}}}$  & 2453 & 2442 & 47.83 & 75.03 & 17.71 & 60.12 \tabularnewline
%\hline
$K_{\mathbf{10}}$  & 1.51 & 7.38 & 8.70 & 14.88 & 8.39 & 14.40 \tabularnewline
%\hline
$K_{\mathbf{\overline{5}}} $  & 5.07 & 6.86 & 14.44 & 11.72 & 14.74 & 0.60 \tabularnewline
%\hline
$m_{\mathbf{5^{\prime}}}$  & 3735 & 3187 & 5.15 & 69.34 & 41.30 & 46.47 \tabularnewline
%\hline
$m_{\mathbf{\overline{5}^{\prime}}}$  & 3780 & 3179 & 64.78 & 14.29 & 88.26 & 17.43 \tabularnewline
%\hline
$a_{\mathbf{5^{\prime}}}$ & -6283 & -4436 & -98.72 & -97.67 & 8.94 & -47.12 \tabularnewline
%\hline
$a_{\mathbf{\overline{5}^{\prime}}}$ & 4606 & -1639 & -88.26 & -1.10 & 22.26 & -10.12 \tabularnewline
%\hline
$ M_{1/2}$ & 944.8 & 781.2 & 1247 & 1249 & 2875 & 1611 \tabularnewline
%\hline
$ \rho_1 $ & 1.00 & 1.00 & 10.00 & 10.00 & 6.14 & 4.96 \tabularnewline
%\hline
$\rho_2$ & 1.00 & 1.00 & 2.00 & 2.00 & 2.12 & 2.47 \tabularnewline
\hline
%\hline \hline
\end{tabular}
     \caption{\it GUT scale parameters for our six benchmark scenarios. Masses and trilinear couplings are in GeV. $M_{1/2}$ is the value of $M_3$ at the GUT scale.}\label{tab:Inputs}
\end{table}

\begin{table}[h!]
\centering
\begin{tabular}{>{\centering}m{1.cm}>{\centering}m{2.cm}>{\centering}m{2.cm}>{\centering}m{2.25cm}>{\centering}m{2.25cm}
>{\centering}m{1.8cm}>{\centering}m{1.8cm}}
 & $\rm{BP1}SU(5)_{\mathbf{1}}$ & $\rm{BP2}SU(5)_{\mathbf{1}}$ &  $\rm{BP1}SU(5)_{\mathbf{200}}$ &  $\rm{BP2}SU(5)_{\mathbf{200}}$ & BPO-I & BPO-II \tabularnewline
\hline \\[-4mm]
%\hline \hline
$m_{h^0}$  & 123.8 & 124.9 & 122.6 & 122.6 & 125.5 & 123.6 \tabularnewline
%\hline
$m_{A^0}$  & 4412 & 3144 & 2592 & 2375 & 3781 & 2635 \tabularnewline
%\hline
$m_{H^0}$  & 4412 & 3144 & 2592 & 2375 & 3781 & 2635 \tabularnewline
%\hline
$m_{H^{\pm}}$  & 4413 & 3145 & 2594 & 2377 & 3782 & 2636 \tabularnewline
\hline
\end{tabular}
     \caption{\it Higgs masses in GeV for our six benchmark scenarios.}\label{tab:Higgs}
\end{table}

\begin{table}[h!]
\centering
\begin{tabular}{>{\centering}m{1.cm}>{\centering}m{2.cm}>{\centering}m{2.cm}>{\centering}m{2.25cm}>{\centering}m{2.25cm}
>{\centering}m{1.8cm}>{\centering}m{1.8cm}}
 & $\rm{BP1}SU(5)_{\mathbf{1}}$ & $\rm{BP2}SU(5)_{\mathbf{1}}$ &  $\rm{BP1}SU(5)_{\mathbf{200}}$ &  $\rm{BP2}SU(5)_{\mathbf{200}}$ & BPO-I & BPO-II \tabularnewline
\hline \\[-4mm]
%\hline \hline
$m_{\tilde{t}_1}$   & 533.5 & 460.8 & 2303 & 2263 & 5039 & 2508 \tabularnewline
%\hline
$m_{\tilde{t}_2}$  & 2572 & 1920 & 3018 & 3028& 5354 & 3386 \tabularnewline
%\hline
$m_{\tilde{b}_1}$  & 2557 & 1900 & 2309& 2268& 4848 & 2890 \tabularnewline
%\hline
$m_{\tilde{b}_2}$  & 2764 & 2453 & 2642 & 2564 & 5332 & 3376 \tabularnewline
%\hline
$m_{\tilde{\tau}_1}$  & 2437 & 2347 & 2704 & 2638 & 4681 & 2795 \tabularnewline
%\hline
$m_{\tilde{\tau}_2}$  & 3277 & 2465 & 4497 & 4418 & 5960 & 2854 \tabularnewline
%\hline
$m_{\tilde{\nu}^3}$   & 2436 & 2347 & 2703 & 2637 & 4680 & 2852 \tabularnewline
\hline  \\[-4mm]
$m_{\tilde{u}_L}$   & 7609 & 7004 & 2877 & 2878 & 6315 & 3852 \tabularnewline
%\hline
$m_{\tilde{u}_R}$  & 7596 & 6998 & 3797 & 3801 & 6551 & 3498 \tabularnewline
%\hline
$m_{\tilde{d}_L}$  & 7610 & 7004 & 2878 & 2879 & 6316 & 3852 \tabularnewline
%\hline
$m_{\tilde{d}_R}$  & 5702 & 6500 & 2751 & 2763 & 5416 & 3071 \tabularnewline
%\hline
$m_{\tilde{e}_L}$  & 5534 & 6403 & 2790 & 2802 & 5012 & 2927 \tabularnewline
%\hline
$m_{\tilde{e}_R}$  & 7449 & 6896 & 4605 & 4610 & 6478 & 2946 \tabularnewline
%\hline
$m_{\tilde{\nu}^1}$ & 5534 & 6402 & 2789 & 2800 & 5011 & 2926 \tabularnewline
\hline
%\hline \hline
\end{tabular}
     \caption{\it First and third generation sfermion masses (we assume the first and second generation sfermions are degenerate) for our six benchmark scenarios.  All the masses are in GeV}
\label{tab:1st3rd}
\end{table}

\begin{table}[h!]
\centering
\begin{tabular}{>{\centering}m{1.cm}>{\centering}m{2.cm}>{\centering}m{2.cm}>{\centering}m{2.25cm}>{\centering}m{2.25cm}
>{\centering}m{1.8cm}>{\centering}m{1.8cm}}
 & $\rm{BP1}SU(5)_{\mathbf{1}}$ & $\rm{BP2}SU(5)_{\mathbf{1}}$ &  $\rm{BP1}SU(5)_{\mathbf{200}}$ &  $\rm{BP2}SU(5)_{\mathbf{200}}$ & BPO-I & BPO-II \tabularnewline
\hline \\[-4mm]
%\hline \hline
$M_{\tilde{g}}$   & 2298 & 1934 & 2760 & 2763 & 5993 & 3476 \tabularnewline
%\hline
$M_{\tilde{\chi}^0_1}$  & 414.9 & 342.0 & 534.9 & 495.6 & 1167 & 689.8 \tabularnewline
%\hline
$M_{\tilde{\chi}^0_2}$  & 805.4 & 663.6 & 538.8 & 499.4 & 1169 & 692.6 \tabularnewline
%\hline
$M_{\tilde{\chi}^0_3}$  & 2319 & 1288 & 2037 & 2041 & 5002 & 3242 \tabularnewline
%\hline
$M_{\tilde{\chi}^0_4}$  & 2320 & 1292 & 5485 & 5496 & 7861 & 3490 \tabularnewline
%\hline
$M_{\tilde{\chi}^{\pm}_1}$  & 805.5 & 663.6 & 536.6 & 497.3 & 1168 & 691.3 \tabularnewline
%\hline
$M_{\tilde{\chi}^{\pm}_2}$  & 2321 & 1293 & 2037 & 2041 & 5002 & 3242 \tabularnewline
\hline
%\hline \hline
\end{tabular}
     \caption{\it Gaugino masses in GeV for our six benchmark scenarios.}
\label{tab:gaugino}
\end{table}

\begin{table}[h!]
\centering
\begin{tabular}{>{\centering}m{1.cm}>{\centering}m{2.cm}>{\centering}m{2.cm}>{\centering}m{2.25cm}>{\centering}m{2.25cm}
>{\centering}m{1.8cm}>{\centering}m{1.8cm}}
 & $\rm{BP1}SU(5)_{\mathbf{1}}$ & $\rm{BP2}SU(5)_{\mathbf{1}}$ &  $\rm{BP1}SU(5)_{\mathbf{200}}$ &  $\rm{BP2}SU(5)_{\mathbf{200}}$ & BPO-I & BPO-II \tabularnewline
\hline \\[-4mm]
%\hline \hline
$\mu$   & 2275 & 1256 & 512.2 & 471.6 & 1094 & 657.5 \tabularnewline
%\hline
$\tan \beta$  & 9.14 & 23.43 & 22.75 & 30.90 & 38.40 & 26.65 \tabularnewline
%\hline
$\Delta$  & 4978 & 2638 & 75.55 & 78.83 & 2.94 & 9.59 \tabularnewline
%\hline
$\Delta_{\mu}$  & 2433 & 750.1 & 141.7 & 119.0 & 646.3 & 232.7 \tabularnewline
%\hline
$\Omega_c h^2$  & $1.01 \times 10^{-1}$ & $3.66 \times 10^{-2}$ & $3.02 \times 10^{-2}$ & $2.59 \times 10^{-2}$ & $1.30\times 10^{-1}$ & $5.01 \times 10^{-2}$ \tabularnewline
%\hline
LSP~type & Bino & Bino & Higgsino & Higgsino & Higgsino & Higgsino \tabularnewline
\hline
\end{tabular}
 \caption{\it The Higgs parameters $\mu$ (in GeV) and $\tan \beta$ for our six benchmark scenarios. Also shown is the fine-tuning $\Delta$ (which does not include fine-tuning in $\mu$ as described in the text), the fine-tuning from $\mu$ alone, the predicted relic density of Dark Matter, and the predominant component of the LSP.}
\label{tab:pars}
\end{table}

The first two benchmarks $\rm{BP1}SU(5)_{\mathbf{1}}$ and $\rm{BP2}SU(5)_{\mathbf{1}}$ have universal gaugino masses consistent with breaking from a singlet of $SU(5)$ ($\rho_1=\rho_2=1$) and only deviate from non-universality for the scalar masses. Although these scenarios have large fine-tuning (as did all the viable universal gaugino scenarios we found) and therefore are not aesthetically pleasing they are still consistent with experimental bounds so should not be dismissed out of hand. 

The next two benchmarks, $\rm{BP1}SU(5)_{\mathbf{200}}$ and $\rm{BP2}SU(5)_{\mathbf{200}}$ are scenarios for which supersymmetry is broken by a gauge-kinetic function in a ${\mathbf{200}}$ dimensional representation of $SU(5)$. This allows non-universal gaugino masses, and in this case the $U(1)$ gaugino is a factor of $10$ heavier than the $SU(3)$ gaugino at the GUT scale. Although the fine-tuning is still sizeable ($\sim 75$ for both scenarios) it is considerably better than for the universal gaugino masses. 

The final two benchmarks are for orbifold {\it inspired} values of gaugino mass ratios. The benchmark BPO-I is inspired by the BIM O-I orbifold model with $n_H+n_{\bar H} = -4$ and $\delta_{GS} = -5$. The benchmark BPO-II is inspired by the BIM O-II orbifold with $\delta_{GS} = -6$. These both have very low (non-$\mu$) fine-tuning. Remarkably BPO-I is also spot on for the relic density of Dark Matter, but unfortunately its spectrum is very heavy and looks beyond the reach of the $14\,{\rm TeV}$ LHC.

%%%%%%%%%%%%%%%%%%%%%%%%%%%%%%%
\section{Discussion and Conclusion} \label{sec:conc}

In this paper we have investigated Grand Unification with $SU(5)$ boundary conditions. In particular we have relaxed some of the more usual restrictions on the GUT scale masses, allowing scalar masses to vary with generation, and have considered scenarios with non-universal gaugino masses. We have checked that our scenarios are consistent with the new observation of a Higgs boson with mass around $125\,{\rm GeV}$, the so far negative direct LHC searches for supersymmetry and the XENON100 direct Dark Matter searches. The scenarios have the correct vacuum structure at low energies and conform with low energy measurements of $b \to s \gamma$, $B_S \to \mu^+ \mu^-$ and $B \to \tau \nu_\tau$, $g-2$ of the muon. Finally we also insist that the scenarios do not produce a Dark Matter relic density above the experimental bounds of the WMAP and Planck satellites. 

We first studied a model of universal gaugino masses but fail to find any solutions with low fine-tuning. This is not surprising since the fine-tuning from $\mu$ alone grows as the square of $\mu$ indicating that a small value of $\mu$ is required if the model is not to be fine-tuned. Unfortunately this is very difficult to achieve while providing a Higgs boson mass heavy enough for the new resonance and we find no solutions for small $\mu$ that do not have to be fine-tuned in one of the other parameters.  We therefore argue that $\mu$ should not be included in our measure of fine-tuning. This is a pragmatic approach and we do no mean to imply that large fine-tuning in $\mu$ is acceptable. However, $\mu$ is already a parameter that is poorly understood with no justification for its phenomenologically required value, and it is possible that $\mu$ has some mechanism of origin that fixes its value in such a way as to avoid the tuning problem. We therefore attempt instead to minimise only the fine-tuning arising from the soft supersymmetry breaking masses. 

However, even with this relaxation, we are still unable to find scenarios with universal gaugino masses that do not have fine-tuning in the soft masses. We therefore turned out attention to the non-universal gaugino masses, initially scanning over all possible ratios. As one might expect we immediately find many more scenarios that conform with the low energy constraints, but although fine-tuning was reduced we still found very few points with acceptable tuning. We examined the cause of this tuning and find that the tunings are greatly reduced for small values of $m_{10}$, $m_{5^\prime}$ and $a_{5^\prime}$ at the GUT scale. This behaviour does not carry over to the fine-tuning with respect to $M_{1/2}$. We therefore ran an ``enhanced'' scan over the non-universal gaugino mass scenarios, this time setting the scalar masses and trilinears to zero at the GUT, and allowing them to gain non-zero values due to a large value of $M_{1/2}$ in the RGEs. Indeed such scenarios with no fine-tuning were suggested many years ago in Ref.~\cite{Hall:1983iz}, where an R-symmetry was imposed to keep the scalar masses zero. This symmetry is then spontaneously broken in the hidden sector and the breaking is transmitted to the visible sector by supergravity. We note that zero or small GUT scale scalar masses generally predict that the squarks and gluinos be of order the same mass. 

Our enhanced scan revealed many scenarios with low ($<10$) fine-tuning in the soft parameters. To achieve the preferred value for the Dark Matter relic density requires $\mu \sim 1\, {\rm TeV}$. Furthermore, we found that all viable scenarios lie on an ellipse in the plane of $\rho_1$ and $\rho_2$ where $\rho_i=M_i/M_3$ at the GUT scale. Since various theories of new physics at the GUT scale make predictions for the gaugino mass ratios, it is interesting to ask where these theories lie on this plane, and by comparison to the ellipse examine whether or not they are likely to give low energy predictions compatible with experiment while maintaining minimal fine-tuning. In particular we examined the breaking of supersymmetry using hidden sector fields in the $\mathbf{24}$, $\mathbf{75}$ or $\mathbf{200}$ representations of $SU(5)$ (the  $\mathbf{1}$ predicts universal gaugino masses), and additionally the Brignole, Ib\'a\~nez and Mu\~noz O-I and O-II orbifold models with various modular weights and Green-Schwarz numbers. It should be stressed that for the orbifold models we only considered the effect on the gaugino masses and disregarded the constraints on the scalar masses. We only found three classes of model that provide viable solutions: supersymmetry breaking using hidden sector fields in a $\mathbf{200}$; the O-I orbifold with $n_H+n_{\bar H}=-4$ and $\delta_{GS}=-5$; and the O-II orbifold with $\delta_{GS}=-6$.

Scans particular to these three models were then performed. All three models turn out to be quite restrictive, predicting particle masses in rather narrow ranges. For example the $SU(5)_\mathbf{200}$ model requires a lightest stop in the region $2.25$ - $2.43\,{\rm TeV}$ and the lightest squark in the region $2.74$ - $2.89\,{\rm TeV}$. Unfortunately the $SU(5)_\mathbf{200}$ model is always quite fine-tuned with $\Delta \gtrsim 75$ and always gives a Dark Matter relic density considerably below the preferred range. The O-I orbifold, on the other hand, is nearly perfect allowing scenarios with  $\Delta < 10$ and always giving the preferred Dark Matter relic density. Unfortunately it also predicts a rather heavy spectrum which will be beyond the search reach of the $14\,{\rm TeV}$ LHC. The O-II orbifold is a half-way house, with an accessible spectrum, scenarios that have low fine-tuning and the possibility for the preferred relic density. Unfortunately the latter two properties are not united in a single scenario, so one must chose between low fine-tuning or the correct relic density. Nevertheless we believe these scenarios are interesting for consideration at future colliders, so we have presented the spectra of some representation benchmark scenarios. 

Of course this by no means exhausts the possible theories of $SU(5)$ grand unification. There are plenty more viable points in the $\rho_1$ - $\rho_2$ plane that could be explored and we encourage model builders to construct models that make predictions for the gaugino mass ratios that lie on our ellipse. It will also be interesting to analyse GUT theories based on other gauge groups (such as $SO(10)$ and $E_6$) with a similar philosophy to see if one can find additional models with desirable properties. This work has shown that supersymmetry with heavy masses can still be quite natural, and it will be exciting to see if such supersymmetric scenarios can be found at the LHC or its predecessor colliders. 

\section*{Acknowledgements}

D.J.M.~acknowledges partial support from the STFC Consolidated Grant ST/G00059X/1. \linebreak[4] A.P.M.\  acknowledges FCT for the grant SFRH/BD/62203/2009 and partial support by the grant PTDC/FIS/116625/2010. A.P.M.\ also acknowledges the members of the Gr@v group at the Physics Department of the University of Aveiro, particularly Marco Sampaio, Carlos Herdeiro and Jo\~ao Rosa for fruitful  discussions and hospitality during visits to Aveiro. A.P.M.\ and D.J.M.\ would also like to thank David Sutherland for his constant support, criticism and fruitful discussions in the realisation of this work, and Chris White for a critical reading of the manuscript.  D.J.M.\ thanks Joydeep Chakrabortty for helpful discussions on gaugino masses, and the Physical Research Laboratory, Ahmedabad for hospitality.

\end{document}